\documentclass[12pt,prd,onecolumn,showpacs,amsmath,amssymb,aps,floats,floatfix]{revtex4-1}

\usepackage[colorlinks=true,urlcolor=blue,anchorcolor=blue,citecolor=blue,filecolor=blue,linkcolor=blue,menucolor=blue,linktocpage=true]{hyperref} 

\usepackage[inline]{enumitem}
\usepackage[multidot]{grffile}  
\usepackage{dcolumn}
\usepackage{bm}
\usepackage{amsmath}
\usepackage{amsfonts}
\usepackage{amssymb}
\usepackage{color}
\usepackage{latexsym}
\usepackage{slashed} 
\usepackage{pstricks}
\usepackage{indentfirst}
\usepackage{mathrsfs}
\usepackage{multirow}
\usepackage{epsfig,psfrag}
\usepackage{subfigure}
\usepackage{mathtools}
\usepackage{setspace} 
\usepackage[utf8]{inputenc} 
\usepackage[scientific-notation=true]{siunitx} 
\usepackage{listings}
\graphicspath{{fig/}}

\newcommand{\tabincell}[2]{\begin{tabular}{@{}#1@{}}#2\end{tabular}}

\makeatother

\allowdisplaybreaks 

\begin{document}
\title{Probing quadruplet scalar dark matter at current and future $pp$ colliders}
\author{Yu-Pan Zeng}
\author{Chengfeng Cai}
\author{Dan-Yang Liu}
\author{Zhao-Huan Yu}\email[]{yuzhaoh5@mail.sysu.edu.cn}
\author{Hong-Hao Zhang}\email[]{zhh98@mail.sysu.edu.cn}
\affiliation{School of Physics, Sun Yat-Sen University, Guangzhou 510275, China}

\begin{abstract}

We investigate a dark matter model involving an inert $\mathrm{SU}(2)_\mathrm{L}$ quadruplet scalar with hypercharge 1/2. After the electroweak symmetry breaking, the dark sector contains one doubly charged, two singly charged, and two neutral scalars. The lighter neutral scalar can be a viable dark matter candidate. Electroweak production of these scalars at the Large Hadron Collider leads to potential signals in the $\text{monojet} + \slashed{E}_\mathrm{T}$ and $\text{soft-leptons} + \text{jets} + \slashed{E}_\mathrm{T}$ channels. We thus derive constraints on the model by reinterpreting recent experimental searches. Based on simulation, we further evaluate the sensitivity at a future 100~TeV $pp$ collider.

\end{abstract}
\maketitle
\tableofcontents

\clearpage

\section{Introduction}
\label{sec:intro}

Among various candidates of particle dark matter (DM), weakly interacting massive particles (WIMPs) seem rather appealing, because they could naturally predict a thermal relic abundance consistent with the observed value~\cite{Bertone:2004pz,Feng:2010gw,Young:2016ala}.
It is straightforward to construct WIMP models by extending the standard model (SM) with new colorless $\mathrm{SU}(2)_\mathrm{L}$ multiplets in the dark sector~\cite{Mahbubani:2005pt,Cirelli:2005uq,Hambye:2009pw,Cohen:2011ec,Cai:2012kt,Earl:2013jsa,AbdusSalam:2013eya,Fischer:2013hwa,Dedes:2014hga,Ostdiek:2015aga,Cai:2015kpa,Tait:2016qbg,Banerjee:2016hsk,Lu:2016dbc,Cai:2016sjz,Cai:2017wdu,Liu:2017gfg,Xiang:2017yfs,Cai:2017fmr,Lopez-Honorez:2017ora,Cai:2018nob,DuttaBanik:2018emv,Gu:2018kmv,Betancur:2018xtj,Kadota:2018lrt,Wang:2018lhk,Filimonova:2018qdc,Chao:2018xwz,Abe:2019wku,Cheng:2019qbd}, which have electroweak interaction strength by definition.
The DM candidate in such models arises from the electrically neutral components of the multiplets.

If the DM candidate is a scalar particle, the minimal extension is to introduce an inert $\mathrm{SU}(2)_\mathrm{L}$ doublet scalar with hypercharge $Y=1/2$, resulting in the inert doublet model (IDM)~\cite{Deshpande:1977rw,Barbieri:2006dq,Gustafsson:2007pc,Cao:2007rm}.
The term ``inert'' means that there exists an unbroken $Z_2$ symmetry that forbids the doublet gaining a nonzero vacuum expectation value (VEV) and directly coupling to SM fermions.
Consequently, if the lightest component of the doublet is one of the electrically neutral components, it would be stable, acting as a WIMP DM candidate.
A next-to-minimal model can be constructed with an inert triplet scalar of $Y=0$ or $Y=1$~\cite{FileviezPerez:2008bj,Araki:2011hm,JosseMichaux:2012wj,Ayazi:2014tha,Khan:2016sxm}.

In this paper, we go further to study a scalar DM model with an inert quadruplet scalar of  $Y=1/2$~\cite{AbdusSalam:2013eya,Cai:2017wdu}, dubbed the quadruplet scalar dark matter (QSDM) model, which has been much less investigated in the past.
The study in Ref.~\cite{AbdusSalam:2013eya} focused on how this model can support a strong first-order electroweak phase transition, as well as the constraints from electroweak oblique parameters, invisible Higgs decay, direct DM detection, and relic abundance.
In our previous work~\cite{Cai:2017wdu}, we investigated the projected sensitivity to this model from improved determination of electroweak oblique parameters in the future Circular Electron-Positron Collider (CEPC) project~\cite{CEPCStudyGroup:2018ghi}.
In this work, we concentrate on production signals of the new scalar bosons in the model at the Large Hadron Collider (LHC) and future $pp$ colliders, which have not been studied in the previous literature.

In the QSDM model, there are three types of independent quartic couplings between the quadruplet and the SM Higgs doublet, which contribute to the mass terms of the quadruplet components due to the nonzero Higgs VEV.
As a result, the components of the quadruplet are split in mass.
Mass eigenstates in the dark sector include two neutral scalars, two singly charged scalars, and one doubly charged scalar.
The lighter neutral scalar could be a viable DM candidate.
DM scattering off nuclei can be mediated by the Higgs boson through the quartic couplings, leading to possible signals in direct detection experiments.

Moreover, the dark sector scalars could be produced in pairs at the LHC via electroweak gauge interactions.
Because of the $Z_2$ symmetry, all these scalars finally decay into the DM particle, which can escape from the LHC detectors, resulting in a large missing transverse energy ($\slashed E_\mathrm{T}$) in the final state.
Since the mass spectrum in the dark sector is typically compressed, visible decay products from the scalars tend to be soft.
Therefore, a hard jet from initial state radiation may be required for triggering the signal at the LHC.
Thus, one possible searching channel is the $\text{monojet} + \slashed E_\mathrm{T}$ channel, which has been widely applied for searching dark matter~\cite{Beltran:2010ww,Rajaraman:2011wf,Fox:2011pm,Yu:2012kj,Xiang:2015lfa,Wang:2017sxx}.
Furthermore, additional soft leptons may contain imprints of the scalar decays~\cite{Giudice:2010wb,Gori:2013ala,Schwaller:2013baa,Han:2014kaa,Baer:2014kya}.
This motivates us to study a $\text{soft-leptons} + \text{jets} + \slashed E_\mathrm{T}$ channel as well.
We will estimate the related constraints on the QSDM model by reinterpreting the existed LHC searches.

At the LHC energies, electroweak production rates for the dark sector scalars are quite low, and, hence, the constraints from current LHC searches are still weak.
Nevertheless, future $pp$ colliders with much higher energies have been proposed, including the Super Proton-Proton Collider (SPPC) at $\sqrt{s} \sim 70\text{--}100~\si{TeV}$~\cite{CEPC-SPPCStudyGroup:2015csa} and the $pp$ Future Circular Collider (FCC-hh) at $\sqrt{s} \sim 100~\si{TeV}$~\cite{Abada:2019lih}.
The increase of the collision energy makes it possible to probe much heavier electroweak scalars.
We thus explore the sensitivity to the QSDM model at a 100~TeV $pp$ collider based on Monte Carlo simulation.

This paper is organized as follows.
In Sec.~\ref{sec:model}, we introduce the model details.
In Sec.~\ref{sec:dmdd}, we identify the parameter regions that are consistent with the observed relic abundance and study the constraints from direct detection experiments.
In Sec.~\ref{sec:monojet}, we explore the constraint from the LHC search in the $\text{monojet} + \slashed E_\mathrm{T}$ channel, as well as the sensitivity at a 100~TeV $pp$ collider.
In Sec.~\ref{sec:soft}, the $\text{soft-leptons} + \text{jets} + \slashed E_\mathrm{T}$ channel is studied.
Section~\ref{sec:con} gives the conclusions and discussions.

\section{Quadruplet scalar dark matter model}
\label{sec:model}

In the QSDM model, we introduce a $\mathrm{SU}(2)_\mathrm{L}$ quadruplet scalar $X$ with hypercharge $Y=1/2$~\cite{AbdusSalam:2013eya,Cai:2017wdu}.
We assume that $X$ is inert; i.e., $X$ is odd under a $Z_2$ symmetry, but all SM fields are $Z_2$ even.
On the one hand, we can express the quadruplet in the vector notation $X=(X^{++},X^+,X^0,X^-)^\mathrm{T}$ with  explicitly indicated electric charges.
On the other hand, it can be denoted by a totally symmetric $\mathrm{SU}(2)_\mathrm{L}$ tensor $X^{ijk}$ ($i,j,k=1,2$).
The components in the two notations are related by
\begin{equation} 
X=\begin{pmatrix}X^{++}\\X^{+}\\X^{0}\\X^{-}\end{pmatrix}=\begin{pmatrix}X^{111}\\ \sqrt{3}X^{112}\\ \sqrt{3}X^{122}\\X^{222}\end{pmatrix}.
\end{equation}
Note that $X^+\neq (X^-)^*$.
The neutral component $X^0$ can be separated into two real scalars $\phi$ and $a$:
\begin{equation}
X^0 =  \frac{1}{\sqrt{2}} (\phi + i a).
\end{equation}

The Lagrangian in the QSDM model is given by
\begin{equation} 
\mathcal{L}=\mathcal{L}_{\mathrm{SM}}+\left(D_{\mu}X\right)^{\dag}D^{\mu}X-V(X),
\end{equation}
where $\mathcal{L}_{\mathrm{SM}}$ is the SM Lagrangian and $V(X)$ is the potential involving $X$.
The covariant derivative for $X$ is $D_{\mu}=\partial_{\mu}-igW_{\mu}^aT^a-i g^{\prime}B_{\mu}/2$, where $T^a$ are the $\mathrm{SU}(2)_\mathrm{L}$ generators in the  representation $\mathbf{4}$.
Electroweak gauge interaction terms for the quadruplet are explicitly given in Appendix~\ref{app:gauge_part}.

Respecting the $Z_2$ symmetry $X^{ijk}\to -X^{ijk}$, we write down the potential $V(X)$ as
\begin{eqnarray}\label{potential}
V(X) &=& M_X^2|X|^2 + \lambda_0 |X|^2 |H|^2
+ \lambda_1 X_{ijk}^{\dag}X^{ijl}H_l^{\dag}H^k
+ \big(\lambda_2 X^{ikl}X^{jmn} H_i^{\dag}H_j^{\dag}\epsilon_{km}\epsilon_{ln} + \mathrm{H.c.}\big)
\nonumber\\
&& +\text{ self-interaction terms of $X$},
\end{eqnarray} 
where $H$ is the SM Higgs doublet.
Here we adopt a convention $\epsilon^{12}=1=-\epsilon_{12}$ for the asymmetric tensors $\epsilon^{ij}$ and $\epsilon_{ij}$.
We do not give the explicit forms for the quadruplet self-interaction terms, because they  will not affect the following discussions.
Note that one may write down an extra operator $X_{ijk}^\dag {X^{ijl}}H_m^\dag {H^n}{\varepsilon ^{km}}{\varepsilon _{ln}}$, but it is not independent, because $X_{ijk}^\dag {X^{ijl}}H_m^\dag {H^n}{\varepsilon ^{km}}{\varepsilon _{ln}} = X_{ijk}^\dag {X^{ijl}}H_l^\dag {H^k} - |X{|^2}|H{|^2}$.
If $\lambda_2$ is complex, we can always make it real by a phase redefinition of the quadruplet.
Hereafter, we just use a real $\lambda_2$.
Since the one-loop contributions to the beta function of the quartic Higgs coupling $\lambda$ from $\lambda_0$, $\lambda_1$, and $\lambda_2$ are all positive~\cite{Hamada:2015bra},
the Higgs vacuum stability problem in the SM~\cite{Degrassi:2012ry} would be partially alleviated in the QSDM model.

After $H$ gets its VEV $v=246.22~\si{GeV}$, mass terms for the quadruplet components can be expressed as 
\begin{equation} 
\mathcal{L}_\mathrm{mass}=-\frac{1}{2}m_{\phi}^2 \phi^2-\frac{1}{2}m_{a}^2 a^2-\begin{pmatrix}\left(X^{+}\right)^*&X^{-}\end{pmatrix}M_\mathrm{C}^2\begin{pmatrix}X^{+}\\\left(X^{-}\right)^*\end{pmatrix}-m_{++}^2|X^{++}|^2,
\end{equation}
with
\begin{eqnarray}
m_{\phi}^2 &=& M_X^2+\frac{1}{6}(3\lambda _0  +2\lambda _1 - 4\lambda _2) v^2,
\label{eq:m_phi}\\
m_a^2 &=& M_X^2+ \frac{1}{6}(3\lambda _0 + 2\lambda _1+ 4\lambda _2) v^2,
\\
M_\mathrm{C}^2 &=&
\begin{pmatrix}
M_X^2+ ( 3\lambda _0+ \lambda _1 )v^2/6 & \lambda _2 v^2/\sqrt{3}\\
\lambda _2 v^2/\sqrt{3} & M_X^2+( \lambda _0+  \lambda _1)v^2/2
\end{pmatrix},
\\
m_{++}^2 &=& M_X^2+\frac{1}{2} \lambda _0 v^2.
\end{eqnarray}
The mass-squared matrix $M_\mathrm{C}^2$ for the singly charged scalars can be diagonalized by a $2\times 2$ rotation matrix $O$, which satisfies
\begin{eqnarray}
O^\mathrm{T} M_\mathrm{C}^2 O &=& \begin{pmatrix}
m_1^2 & \\
 & m_2^2
\end{pmatrix},
\\
O &=& \begin{pmatrix}
\cos\theta & -\sin\theta\\
\sin\theta & \cos\theta
\end{pmatrix}.
\end{eqnarray}
The rotation angel $\theta$ can be obtained from
\begin{equation}
\sin\theta = \frac{{ - \sqrt 6 {\lambda _2}}}{{\sqrt {\lambda _1^2 + 12\lambda _2^2 + {\lambda _1}\sqrt {\lambda _1^2 + 12\lambda _2^2} } }}.
\end{equation}
Thus, the singly charged mass eigenstates $X^+_1$ and $X^+_2$ are related to the gauge eigenstates $X^+$ and $(X^-)^*$ through
\begin{equation}
\begin{pmatrix}
   {{X^ + }}  \\
   {{{({X^ - })}^*}}  \\
\end{pmatrix}
= O\begin{pmatrix}
   {X_1^ + }  \\
   {X_2^ + }  \\
\end{pmatrix}.
\end{equation}
Their masses squared are given by
\begin{eqnarray}
m_1^2 &=& M_X^2 + \frac{{{v^2}}}{{12}}\Big( {6{\lambda _0} + 4{\lambda _1} - 2\sqrt {\lambda _1^2 + 12\lambda _2^2} \,} \Big),
\\
 m_2^2 &=& M_X^2 + \frac{{{v^2}}}{{12}}\Big( {6{\lambda _0} + 4{\lambda _1} + 2\sqrt {\lambda _1^2 + 12\lambda _2^2} \,} \Big).
\end{eqnarray}

The mass hierarchy of the neutral scalars $\phi$ and $a$ is determined by the sign of $\lambda_2$.
If $\lambda _2 > 0$ ($\lambda _2 < 0$), $\phi$ is lighter (heavier) than $a$, and, thus, $\phi$ ($a$) is a possible DM candidate.
Nevertheless, if $|{\lambda _1}| > 2|{\lambda _2}|$, one of the singly charged scalars is lighter than the DM candidate.
Additionally, if ${\lambda _1} > 2|{\lambda _2}|$, the doubly charged scalar is lighter than the DM candidate.
Since the DM candidate should be the lightest particle in the dark sector for ensuring its stability, we have the following conclusions.
\begin{itemize}
\item If $\lambda_2 >0$ and $|\lambda_1| \leq 2 \lambda_2$, then $\phi$ is a viable DM candidate.
\item If $\lambda_2 <0$ and $|\lambda_1| \leq -2 \lambda_2$, then $a$ is a viable DM candidate.
\end{itemize}

Similar to the IDM, the QSDM model has two kinds of $CP$ symmetries, one with $\phi\to \phi$ and $a\to -a$ and the other one with $\phi\to -\phi$ and $a\to a$~\cite{Belyaev:2016lok}.
A transformation $X^{ijk}\to iX^{ijk}$ and $\lambda_2 \to -\lambda_2$ can keep the Lagrangian unchanged but interchange the two $CP$ symmetries and, hence, the roles of $\phi$ and $a$.
Therefore, we know that $\phi$ and $a$ have opposite $CP$ parities, but it is impossible to determine their absolute $CP$ parities without additional interactions.
Without loss of generality, hereafter we adopt $\lambda_2>0$ and take $\phi$ as the DM candidate.
The resulting discussions are totally equivalent to those for $\lambda_2<0$ and $a$ as the DM candidate.

In the following analyses, four free parameters in the QSDM model are chosen to be $\{M_X, \lambda_0, \lambda_1, \lambda_2\}$.
The parameter space is analogous to that of the IDM~(cf. Refs.~\cite{Barbieri:2006dq,Gustafsson:2007pc,Cao:2007rm,Belyaev:2016lok}) in the sense of the number and the roles of the parameters.
Nonetheless, the number of dark sector scalars in the QSDM model is more.
The IDM dark sector includes two neutral scalars with opposite $CP$ and one singly charged scalar.
The neutral scalars in the two models play similar roles, with the lighter one being the DM candidate.
On the other hand, the QSDM model contains one more singly charged scalar and an additional doubly charged scalar.
After electroweak symmetry breaking, the mass eigenstates of singly charged scalars are different from the gauge eigenstates.
This is a new phenomenon that does not exhibit in the IDM.

\section{Relic abundance and direct detection}
\label{sec:dmdd}

In this section, we evaluate the relic abundance prediction in the QSDM model, and investigate the constraints from direct detection experiments.

The dark sector scalars can interact with SM particles via electroweak gauge couplings and scalar couplings to the Higgs boson.
Through such interactions, these scalars could be thermally produced in the early Universe and decoupled from the cosmic plasma at the freeze-out epoch.
Conventionally, the relic abundance of dark matter is determined by its freeze-out annihilation cross section.
Nonetheless, for $m_X\sim \mathcal{O}(\si{TeV})$, the mass splittings among the dark sector scalars due to the quartic couplings would be relatively small, and, thus, the scalars actually freeze out around the same epoch.
Therefore, the coannihilation effect would be significant for evaluating the relic abundance~\cite{Griest:1990kh}.

There are a lot of relevant annihilation and coannihilation processes.
For instance, a $\phi \phi$ pair can annihilate into a SM fermion pair $f\bar{f}$, or an electroweak gauge boson pair $W^+W^-$ or $ZZ$, or a Higgs boson pair $hh$.
Some of these annihilation processes are mediated by $s$-channel $Z$ and Higgs bosons, while the others are related to the exchanges of $t$- and $u$-channel dark sector scalars as well as to the quartic couplings.
Because of the significant coannihilation effect, it is not sufficient to just consider the processes that are directly related to DM annihilation.
Actually, annihilation or coannihilation between every pair of dark sector scalars could affect the final DM relic abundance.

We utilize a few numerical tools to predict the relic abundance of the DM candidate $\phi$.
\texttt{FeynRules~2}~\cite{Alloul:2013bka} is adopted to implement the QSDM model, interfaced to the Monte Carlo generator \texttt{MadGraph5\_aMC@NLO~2}~\cite{Alwall:2014hca}.
The relic abundance $\Omega_\phi h^2$ is calculated by a \texttt{MadGraph} plugin \texttt{MadDM}~\cite{Ambrogi:2018jqj}, which can reliably take into account the coannihilation effect.
All annihilation and coannihilation diagrams are automatically involved in the calculation.

\begin{figure}[!t]
\centering
\subfigure[~$\lambda_0=0.5$ and $\lambda_1=0.05$ fixed\label{DD_RD:lamb2}]
{\includegraphics[width=.49\textwidth]{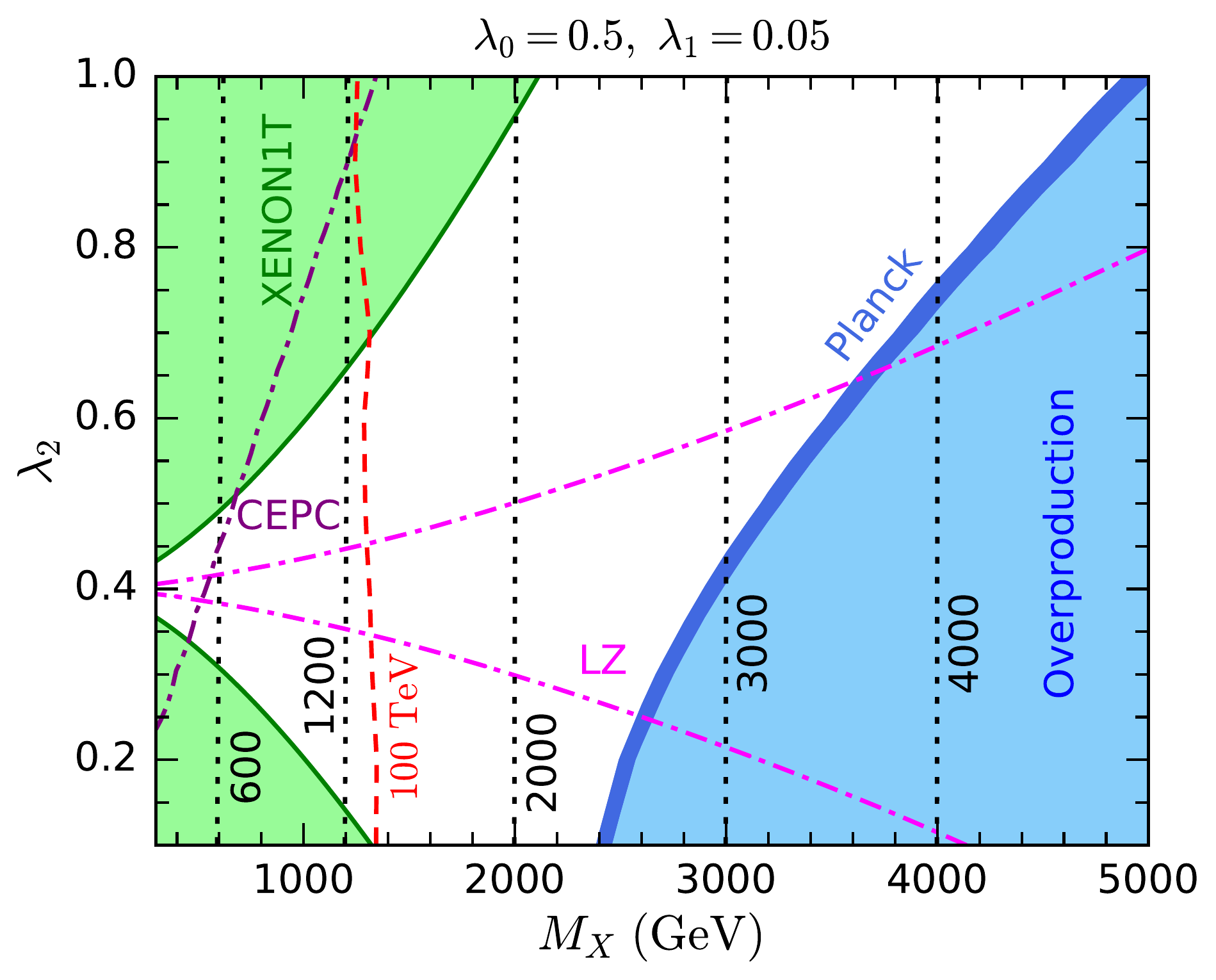}}
\subfigure[~$\lambda_0=0.7$ and $\lambda_2=0.5$ fixed\label{DD_RD:lamb1}]
{\includegraphics[width=.49\textwidth]{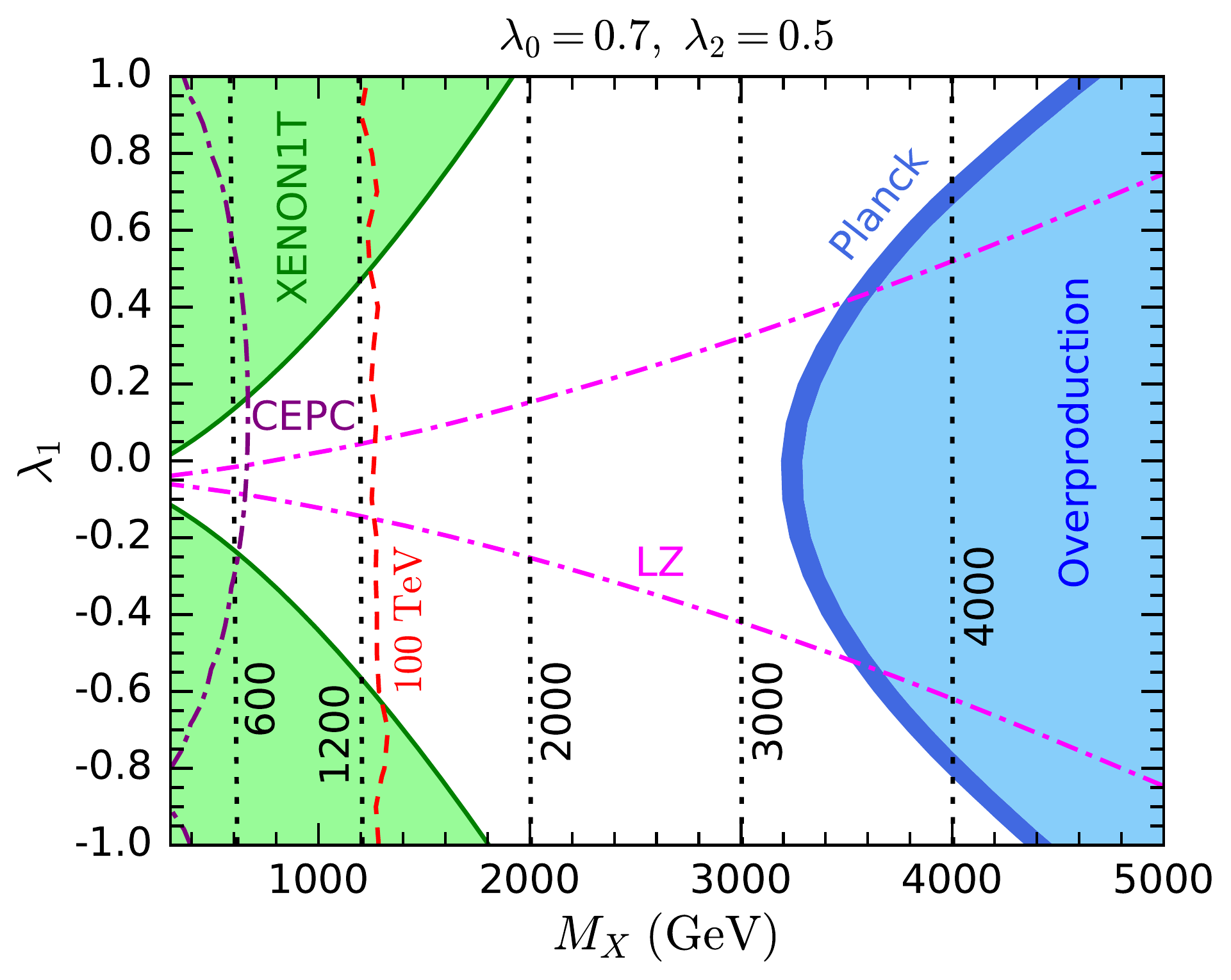}}
\caption{Experimental constraints and sensitivities in the $M_X$-$\lambda_2$ (a) and $M_X$-$\lambda_1$ (b) planes.
The black dotted lines denote the DM candidate mass $m_\phi$ in GeV.
The blue bands correspond to the $3\sigma$ range of the Planck relic abundance measurement~\cite{Aghanim:2018eyx},
while the light blue regions predict overproduction of dark matter.
The green regions are excluded by the direct detection experiment XENON1T~\cite{Aprile:2018dbl}.
The dot-dashed magenta lines indicate the exclusion capability of the future direct detect experiment LZ~\cite{Mount:2017qzi}.
The dot-dashed purple lines show the expected sensitivities of the measurement of electroweak oblique parameters at the future CEPC.
The dashed red lines show the exclusion capability of the $\text{monojet} + \slashed E_\mathrm{T}$ channel at a 100~TeV $pp$ collider with a dataset of $3~\si{ab^{-1}}$ (see Sec.~\ref{sec:monojet}).
}
\label{DD_RD}
\end{figure}

The measurement of the DM relic abundance in the Planck experiment gives $\Omega_\mathrm{DM} = 0.1200 \pm 0.0012$~\cite{Aghanim:2018eyx}.
In Figs.~\ref{DD_RD:lamb2} and \ref{DD_RD:lamb1}, we fix the parameters $(\lambda_0,\lambda_1)=(0.5,0.05)$  and $(\lambda_0, \lambda_2)=(0.7,0.5)$ and show the parameter regions that are consistent with the Planck observation as the blue bands in the $M_X$-$\lambda_2$ and $M_X$-$\lambda_1$ planes, respectively.
The black dotted lines indicate the contours of the DM candidate mass $m_\phi$, which slightly deviates from $M_X$ due to the quartic couplings.

If $M_X$ increases, the effective annihilation cross section typically decreases, leading to an increase in the relic abundance.
Therefore, the light blue regions with large $M_X$ predict overproduction of $\phi$ particles in the early Universe, which contradicts standard cosmology.
For small values of $\lambda_2$ ($|\lambda_1|$) in Fig.~\ref{DD_RD:lamb2} [Fig.~\ref{DD_RD:lamb1}], the relic abundance observation corresponds to $M_X \sim 2.4 ~(3.3)~\si{TeV}$, which increases to $M_X \sim 5~(4.6)~\si{TeV}$ when $\lambda_2$ ($|\lambda_1|$) increases to one.
These results are consistent with the simplified calculation given in Ref.~\cite{Cirelli:2005uq}.

Direct detection experiments look for signals of DM scattering off nuclei.
In the QSDM model, DM scattering is mediated by the Higgs boson $h$, arising from the quartic potential terms that lead to the $h\phi \phi$ interaction Lagrangian
\begin{eqnarray}
{\mathcal{L}_{h\phi \phi }} &=& \frac{1}{2}{\lambda _{h\phi \phi }}vh{\phi ^2},
\\
{\lambda _{h\phi \phi }} &=&  - {\lambda _0} - \frac{2}{3}{\lambda _1} + \frac{4}{3}{\lambda _2}.
\label{eq:lam_hphiphi}
\end{eqnarray}
As direct detection experiments basically operate at zero momentum transfer, the interactions between DM and quarks can be described by dimension-5 effective operators~\cite{Yu:2011by}.
As a result, the spin-independent DM-nucleon scattering cross section can be expressed as
\begin{equation}
\sigma _{\chi N}^{{\mathrm{SI}}} = \frac{{m_N^2F_N^2}}{{4\pi {{({m_\phi } + {m_N})}^2}}},\quad
N=p,n,
\end{equation}
where
\begin{equation}
{F_N} =  - \frac{{{\lambda _{h\phi \phi }}{m_N}}}{{9m_h^2}}[2 + 7(f_u^N + f_d^N + f_s^N)].
\end{equation}
Here the nucleon form factors $f_q^N$ are given by~\cite{Ellis:2000ds}
\begin{eqnarray}
&&f_u^p=0.020\pm 0.004,~~ f_d^p=0.026\pm 0.005,~~ f_u^n=0.014\pm 0.003,\nonumber\\
&&f_d^n=0.036\pm 0.008,~~ f_s^p=f_s^n=0.118\pm 0.062 .~~
\end{eqnarray}

In Figs.~\ref{DD_RD:lamb2} and \ref{DD_RD:lamb1}, we show the parameter regions excluded by the direct detection experiment XENON1T~\cite{Aprile:2018dbl} at 90\% confidence level (C.L.).
According to Eq.~\eqref{eq:lam_hphiphi}, we can take some particular relations among $\lambda_0$, $\lambda_1$, and $\lambda_2$ to give a vanishing $h\phi\phi$ coupling, resulting in ``blind spots'' for direct detection experiments.
These relations correspond to the flat directions
among the scalar couplings, where the Higgs VEV has zero contribution to $m_\phi$.
For Figs.~\ref{DD_RD:lamb2} and \ref{DD_RD:lamb1},
the limits $\lambda_2 = 3{\lambda _0}/4 + {\lambda _1}/2 = 0.4$ and $\lambda_1 = - 3{\lambda _0}/2 + 2{\lambda _2} =-0.05$ correspond to $\lambda_{h\phi \phi }=0$, respectively.
Therefore, direct detection experiments lose their sensitivities as $\lambda_2$ or $\lambda_1$ approaches the corresponding limit.
Nonetheless, the XENON1T experiment has excluded some disconnected parameter regions with $M_X \lesssim 1.3\text{--}2~\si{TeV}$.
We also demonstrate the expected 90\% C.L. exclusion limits of the future direct detection experiment LZ~\cite{Mount:2017qzi}, which will explore the parameter space much deeper and be able to reach the regions suggested by the relic abundance measurement.

Note that the results presented here are based on tree-level calculations.
There are also contributions from electroweak loop-induced diagrams~\cite{Cirelli:2005uq,Dedes:2014hga}, leading to a nonvanishing spin-independent cross section for the tree-level blind spots.
Nevertheless, one would expect a cancellation between the tree and loop diagrams if the scalar couplings are carefully tuned.
This means that the blind spots would still exist at loop level, but their positions in the parameter space would be slightly shifted.

As studied in previous papers~\cite{AbdusSalam:2013eya,Cai:2017wdu}, the dark sector scalars in the QSDM model can contribute to the electroweak oblique parameters $S$, $T$, and $U$ at one-loop level, and, hence, affect electroweak precision measurements.
In Fig.~\ref{DD_RD}, we also show  the 95\% C.L. expected sensitivities of the measurement of electroweak oblique parameters at the future CEPC project~\cite{CEPCStudyGroup:2018ghi}.
This result is estimated following the strategy in our previous work~\cite{Cai:2017wdu} with the optimistic settings.
We can see that the CEPC experiment would probe up to $m_X \sim 600 \text{--} 1200~\si{GeV}$, covering some regions related to the blind spots in direct detection.

\section{Monojet searches at $pp$ colliders}
\label{sec:monojet}

\begin{figure}[!t]
\centering
\subfigure[~$W^+$ mediation]
{\includegraphics[height=.2\textwidth]{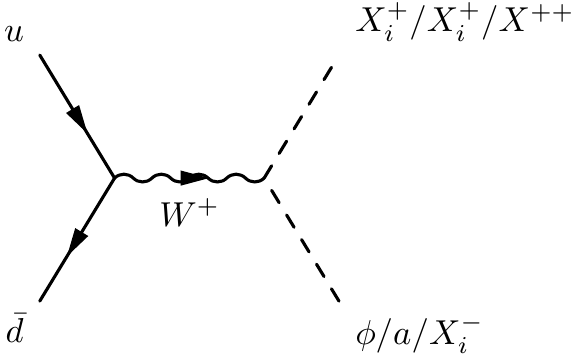}}
\hspace{1em}
\subfigure[~$Z$ mediation]
{\includegraphics[height=.2\textwidth]{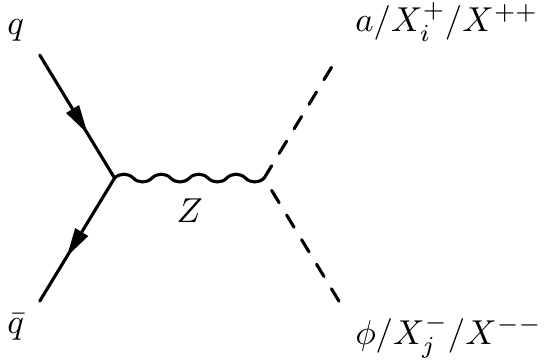}}
\hspace{1em}
\subfigure[~$\gamma$ mediation]
{\includegraphics[height=.2\textwidth]{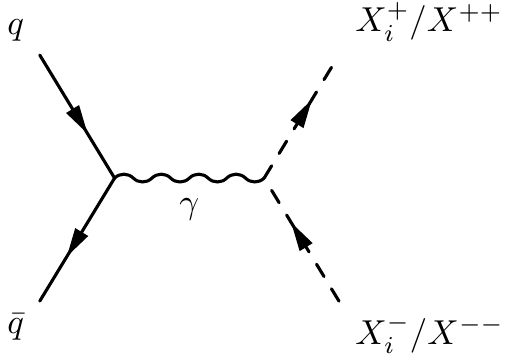}}
\caption{Typical diagrams for pair production of dark sector scalars at parton level in $pp$ collisions, including $W^+$-mediated $u + \bar{d} \to X_i^+/X_i^+/X^{++} + \phi/a/X_i^-$~(a), $Z$-mediated $q + \bar{q} \to a/X_i^+/X^{++} + \phi/X_j^-/X^{--}$~(b), and $\gamma$-mediated $q + \bar{q} \to X_i^+/X^{++} + X_i^-/X^{--}$~(c).
}
\label{prod_diagrams}
\end{figure}

\begin{figure}[!t]
\centering
\subfigure[~$X_i^+/X^{++} \to W^{+(*)} + \phi/X_1^+$]
{\includegraphics[height=.2\textwidth]{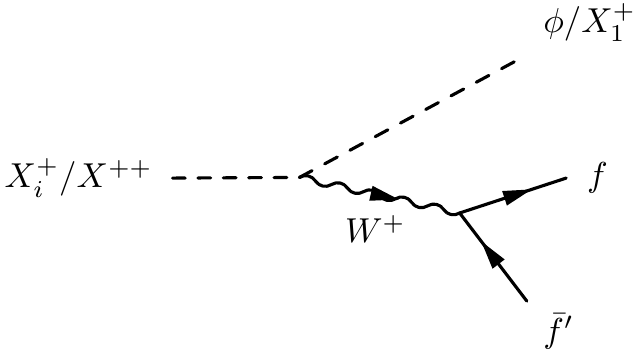}}
\hspace{2em}
\subfigure[~$a/X_2^+ \to W^{-(*)} + X_i^+/X^{++}$]
{\includegraphics[height=.2\textwidth]{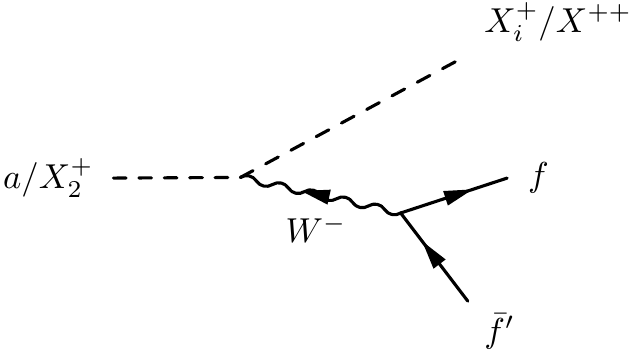}}

\subfigure[~$a/X_2^+ \to Z^{(*)} + \phi/X_1^+$]
{\includegraphics[height=.2\textwidth]{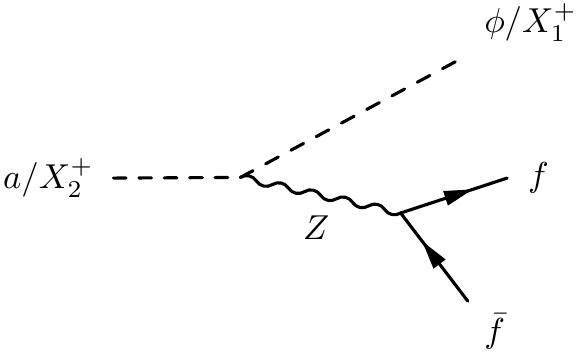}}
\hspace{2em}
\subfigure[~$X_2^+ \to h^{(*)} + X_1^+$]
{\includegraphics[height=.2\textwidth]{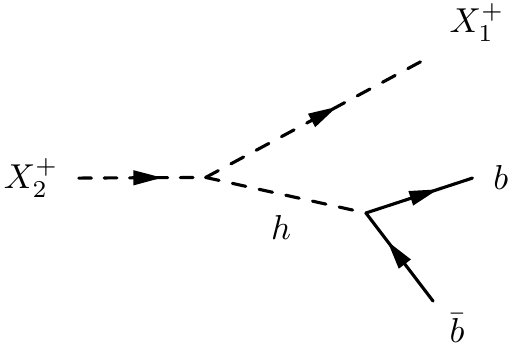}}
\caption{Typical decay diagrams for dark sector scalars, including $X_i^+/X^{++} \to W^{+(*)} + \phi/X_1^+$~(a), $a/X_2^+ \to W^{-(*)} + X_i^+/X^{++}$~(b), $a/X_2^+ \to Z^{(*)} + \phi/X_1^+$~(c), and $X_2^+ \to h^{(*)} + X_1^+$~(d).
}
\label{decay_diagrams}
\end{figure}

Through the electroweak gauge couplings, the dark sector scalars in the QSDM model could be directly produced in pairs at the LHC.
The corresponding processes can be expressed as $pp\to \chi_i\chi_j +\text{jets}$ with $\chi_i = (\phi,a,X_1^\pm,X_2^\pm,X^{\pm\pm})$.
Figure~\ref{prod_diagrams} shows some typical parton-level diagrams for pair production of dark sector scalars at the LHC.
After production, a heavier scalar $\chi_k$ may decay into a lighter scalar $\chi_l$ via $\chi_k\to W^{\pm(*)}/Z^{(*)}/h^{(*)} + \chi_l$.
Typical decay diagrams are demonstrated in Fig.~\ref{decay_diagrams}.
Depending on the mass splitting between $\chi_k$ and $\chi_l$, the produced $W^\pm$, $Z$, and $h$ bosons can be either on or off shell.
Subsequent decays may happen and form decay chains.
Finally, all $Z_2$-odd scalars will decay into the DM candidate $\phi$, which is stable and escapes from detection, leading to a large $\slashed E_\mathrm{T}$.

\begin{figure}[!t]
\centering
\includegraphics[width=.49\textwidth]{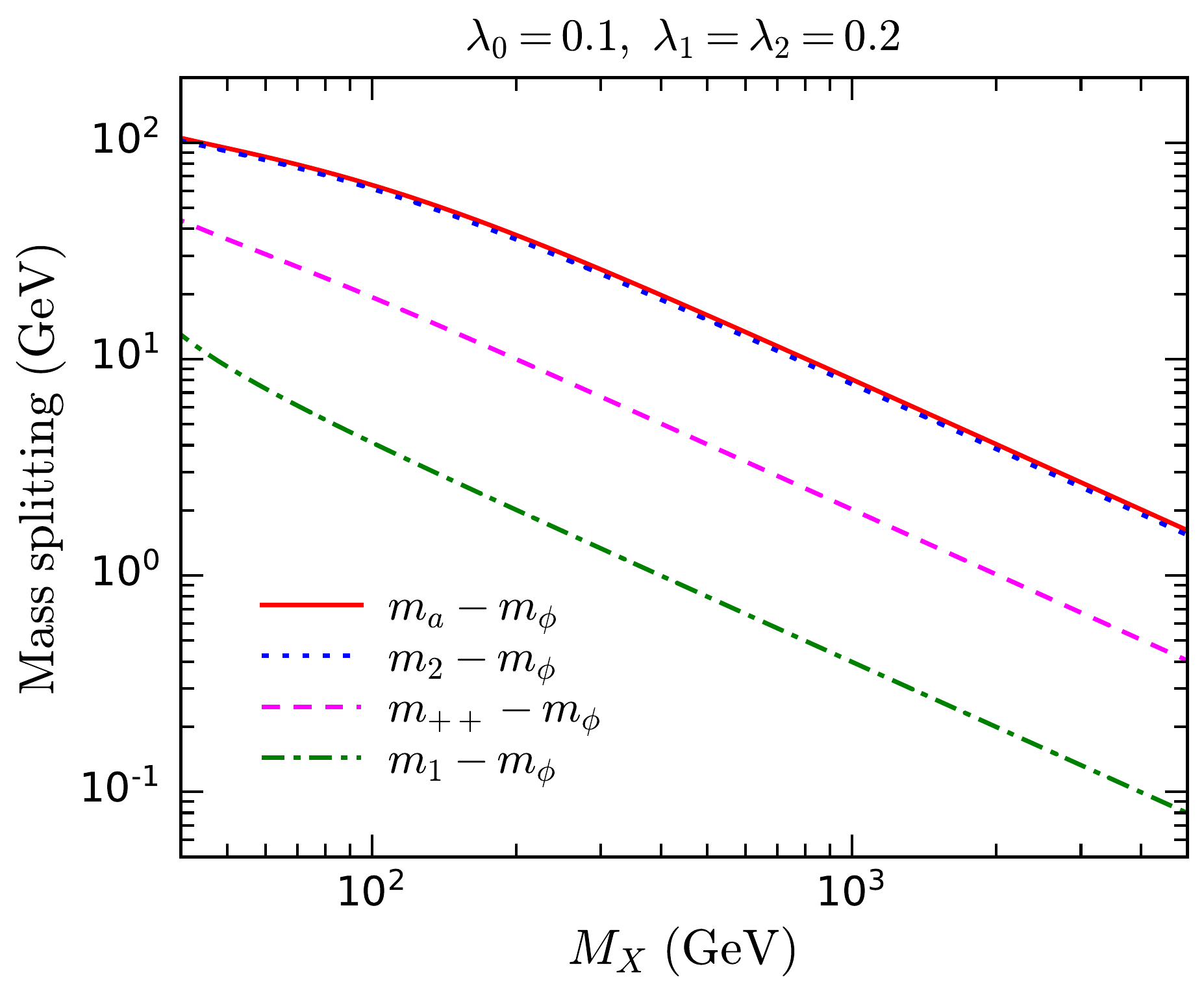}
\caption{Mass splittings as functions of $M_X$ for $\lambda_0 = 0.1$ and $\lambda_1 = \lambda_2 = 0.2$.}
\label{Mass_splitting}
\end{figure}

Figure~\ref{Mass_splitting} shows the mass splittings between $(a,X_1^\pm,X_2^\pm,X^{\pm\pm})$ and $\phi$ as functions of $M_X$ for $\lambda_0 = 0.1$ and $\lambda_1 = \lambda_2 = 0.2$.
From this plot, we can read off the mass difference between each pair of dark sector scalars.
As $M_X$ increases, the contributions from the quartic couplings relatively decrease, resulting in smaller splittings.
The mass splitting between the two neutral scalars $a$ and $\phi$ is the largest one, ranging from $\sim 100$ to $\sim 2~\si{GeV}$ as $m_X$ increases from 40~GeV to 5~TeV.
For $m_X \gtrsim 70~\si{GeV}$, the splittings are not large enough to induce on-shell $W^\pm$, $Z$, or $h$ bosons.
For fixed $M_X$, smaller quartic couplings would further compress the mass spectrum.

In the above analysis, we find that the mass splittings in the QSDM model are typically small.
Consequently, visible decay products from the dark sector scalars would be quite soft and, hence, difficult to be triggered in detectors.
In order to effectively trigger the signal, we can require at least one hard jet from initial state radiation to recoil the $\chi_i\chi_j$ pair, leading to a $\text{monojet}+\slashed E_\mathrm{T}$ final state~\cite{Beltran:2010ww,Rajaraman:2011wf,Fox:2011pm}.
SM backgrounds in the $\text{monojet}+\slashed E_\mathrm{T}$ search channel include two major backgrounds---$W\left(\rightarrow l\nu\right)+\text{jets}$ and $Z\left(\rightarrow \nu\bar{\nu}\right)+\text{jets}$---and some minor backgrounds, such as $t\bar{t}+\text{jets}$ and $VV+\text{jets}$ ($V=W^\pm,Z$).
In these backgrounds, $\slashed E_\mathrm{T}$ mainly arises from neutrinos in the decay products.

\subsection{LHC constraint}

In this subsection, we investigate the current LHC constraint on the QSDM model by reinterpreting the ATLAS analysis in the $\text{monojet}+\slashed E_\mathrm{T}$ channel with an integrated luminosity of $36.1~\si{fb^{-1}}$ at  $\sqrt{s}=13~\mathrm{TeV}$~\cite{Aaboud:2017phn}.
For this purpose, we utilize \texttt{MadGraph}~\cite{Alwall:2014hca} to generate signal simulation samples.
Parton shower is performed by \texttt{PYTHIA~8}~\cite{Sjostrand:2014zea} with the MLM matching scheme~\cite{Mangano:2006rw}.
\texttt{PYTHIA} is also carried out for hadronization and decay processes.
Then we use \texttt{Delphes~3}~\cite{deFavereau:2013fsa} for a fast detector simulation with a setup for the ATLAS detector.

We simulate the signal processes $pp\to \chi_i\chi_j +\text{jets}$ and apply the same selection cuts in the ATLAS analysis~\cite{Aaboud:2017phn} to the simulation events.
Isolated leptons, including electrons and muons, and jets are reconstructed with the conditions on $p_\mathrm{T}$ and $\eta$ listed in Table~\ref{t_monojet}.
Then the events in the signal regions are required to have a hard leading jet with $p_\mathrm{T} > 250~\si{GeV}$ and $|\eta| < 2.4$ and a missing transverse energy $\slashed E_\mathrm{T}$ at least larger than $250~\si{GeV}$.
In addition, there should be no leptons and no more than four jets.
Moreover, the separation in the azimuthal angle between any reconstructed jet $j_i$ and the missing transverse momentum $\slashed{\mathbf{p}}_\mathrm{T}$ should satisfy $\Delta\phi(j_i,\slashed{\mathbf{p}}_\mathrm{T}) > 0.4$ for preventing a large $\slashed E_\mathrm{T}$ from mismeasurement of jets.
Finally, ten inclusive and ten exclusive signal regions are defined with different $\slashed E_\mathrm{T}$ thresholds, whose explicit definitions can be found in Table~1 of Ref.~\cite{Aaboud:2017phn}.
In Table~\ref{t_monojet}, we summarize the cut conditions above.

\begin{table}[!t]
\caption{Reconstruction and cut conditions in the $\text{monojet} + \slashed E_\mathrm{T}$ channel.}
\label{t_monojet}
\setlength{\tabcolsep}{1em}
\renewcommand{\arraystretch}{1.2}
\begin{tabular}{ccc}
\hline\hline
 & 13~TeV LHC & 100~TeV $pp$ collider \\
\hline
\multicolumn{3}{c}{Reconstruction conditions}\\
Electron $p_\mathrm{T}$, $|\eta| $&$>20~\mathrm{GeV}$, $<2.47$&$>40~\mathrm{GeV}$, $<2.47$\\
Muon $p_\mathrm{T}$, $|\eta| $&$>10~\mathrm{GeV}$, $<2.5$&$>20~\mathrm{GeV}$, $<2.5$\\
Jet $p_\mathrm{T}$, $|\eta| $&$>30~\mathrm{GeV}$, $<2.8$ & $>60~\mathrm{GeV}$, $<2.8$\\
\hline
\multicolumn{3}{c}{Cut conditions}\\
Number of leptons &$0  $&$0 $\\
Leading jet $p_\mathrm{T}$, $|\eta| $&$>250~\mathrm{GeV},\ <2.4$&$>1.4~\mathrm{TeV}$, $<2.4$\\
Number of jets &$\leq4  $&$\leq4 $\\
$\Delta\phi(j_i,\slashed{\mathbf{p}}_\mathrm{T})$ & $>0.4$ & $>0.4$\\
$\slashed E_\mathrm{T} $ & $>250\text{--}1000~\mathrm{GeV} $&$>1.5\text{--}2.8~\mathrm{TeV}$\\
\hline\hline
\end{tabular}
\end{table}

\begin{figure}[!t]
\centering
\subfigure[~$\lambda_1=\lambda_2 = 3\lambda_0/2$\label{monoj0hxx}]
{\includegraphics[width=.49\textwidth]{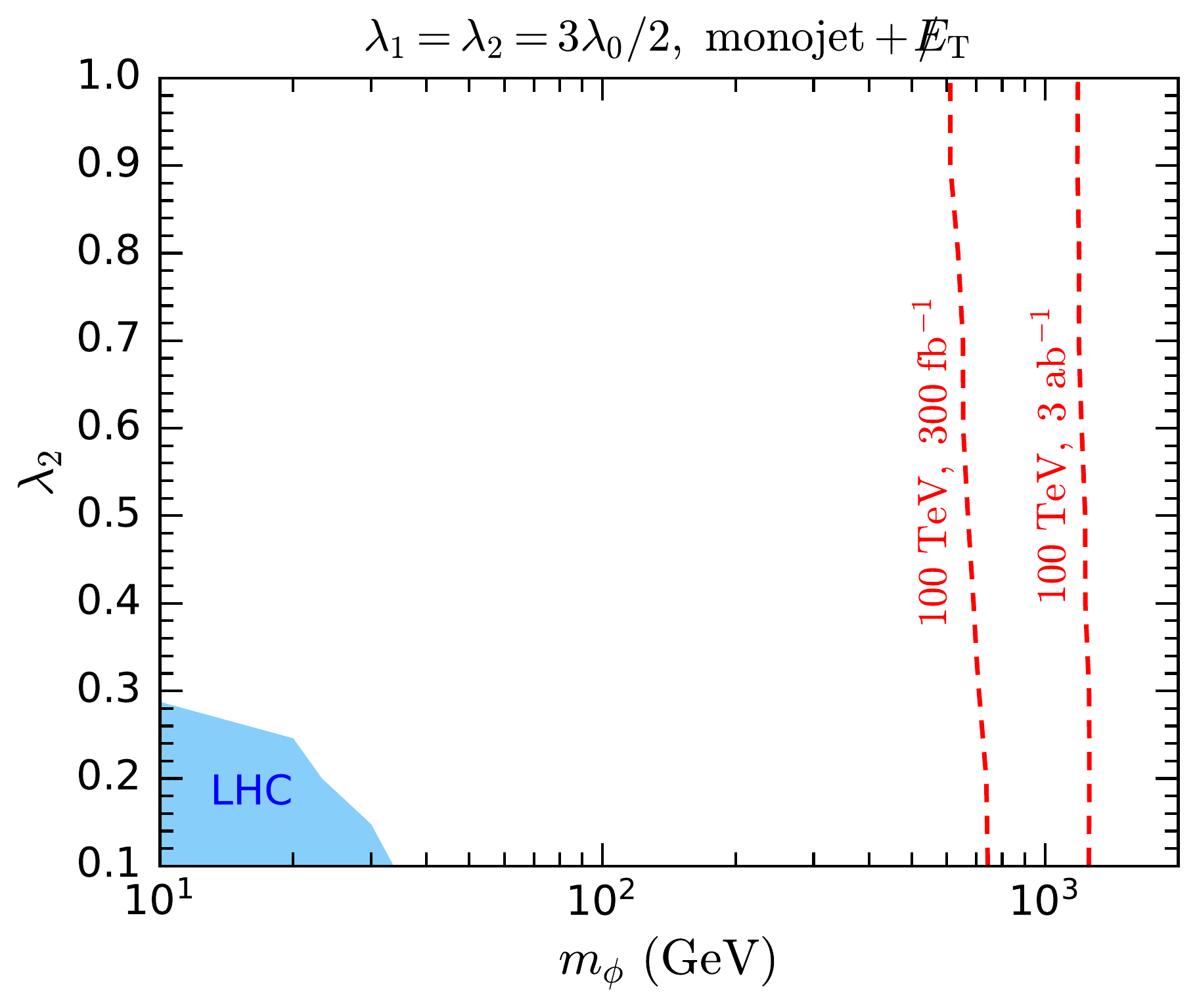}}
\subfigure[~$\lambda_0=0$, $\lambda_1=2\lambda_2$~\label{monojjian0hxx}]
{\includegraphics[width=.49\textwidth]{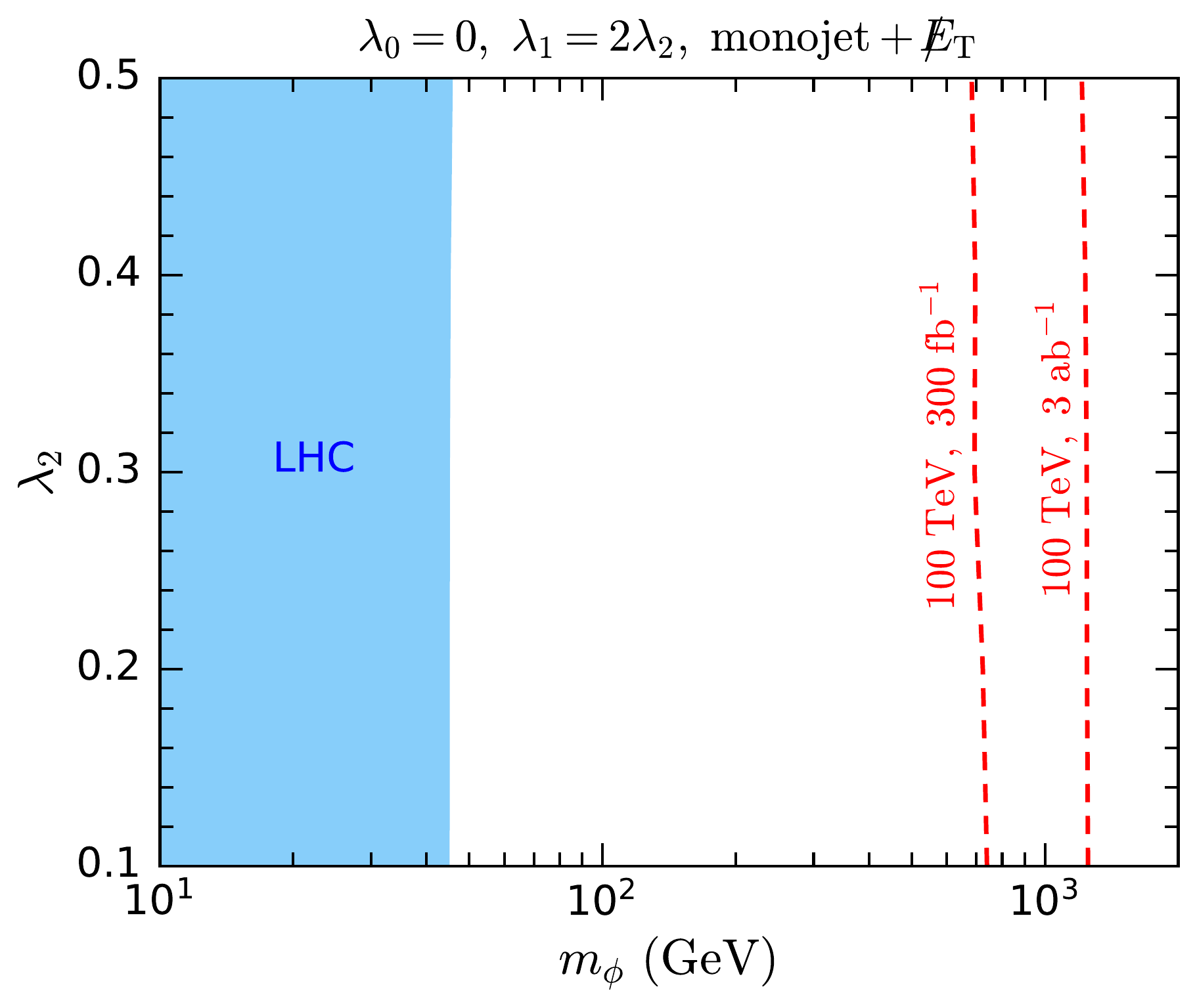}}
\caption{Current constraints and future sensitivities from the $\text{monojet} + \slashed E_\mathrm{T}$ channel in the $m_\phi$-$\lambda_2$ plane for the fixed coupling relations of $\lambda_1=\lambda_2 = 3\lambda_0/2$ (a) and of $\lambda_0=0$ and $\lambda_1=2\lambda_2$ (b).
The blue regions are excluded at 95\% C.L. by the ATLAS search with a $36.1~\si{fb^{-1}}$ dataset at the 13~TeV LHC~\cite{Aaboud:2017phn}.
The red dashed lines denote the 95\% C.L. expected exclusion limits at a 100~TeV $pp$ collider with integrated luminosities of $300~\si{fb^{-1}}$ and $3~\si{ab^{-1}}$.}
\label{jian&0hxx}
\end{figure}

Based on the signal simulation samples, we estimate the visible cross section in each signal region, which is a product of production cross section, acceptance, and efficiency, and then use the 95\% C.L. observed experimental upper limit to derive constraints on the QSDM model.
Taking into account all the signal regions, the combined exclusion region in the $m_\phi$-$\lambda_2$ plane is shown in Fig.~\ref{monoj0hxx}, where we fix a coupling relation of $\lambda_1=\lambda_2 = 3\lambda_0/2$.
Because of Eq.~\eqref{eq:lam_hphiphi}, such a relation leads to $\lambda_{h\phi\phi}=0$, and there is no constraint from direct detection experiments.
Therefore, collider searches are really important in this case.
Note that $\lambda_{h\phi\phi}=0$ also leads to $m_\phi = m_X$, according to Eqs.~\eqref{eq:m_phi} and \eqref{eq:lam_hphiphi}.
We find that the monojet search has excluded a region with $m_\phi \lesssim 33~\si{GeV}$ and $\lambda_2\lesssim 0.3$.
Nonetheless, the sensitivity decreases as $\lambda_2$ increases.
The reason is that a larger $\lambda_2$ leads to larger mass splittings among the dark sector scalars and, hence,1 harder leptons from scalar decays that would not be easy to pass the cuts.

In Fig.~\ref{monojjian0hxx}, we adopt another relation $\lambda_0=0$ and $\lambda_1=2\lambda_2$, which also results in $\lambda_{h\phi\phi}=0$.
Additionally, it leads to degenerate mass spectra with $m_\phi=m_1=m_{++}$ and $m_a = m_2$.
Consequently, many decay channels are turned off, significantly reducing the probability of finding leptons in the final state.
Therefore, the monojet search is more sensitive, excluding a region up to $m_\phi \sim 45~\si{GeV}$.
The exclusion is basically regardless of the $\lambda_2$ value.

\subsection{Sensitivity at a 100 TeV $pp$ collider}

The above results have shown that the current LHC monojet search is rather insensitive to the QSDM model, just probing a scale of a few tens of GeV.
Since production cross sections in $pp$ collisions typically increase as $\sqrt{s}$ increases,
we expect that monojet searches at a future $pp$ collider with $\sqrt{s} \sim 100~\si{TeV}$ would be much more sensitive.
Below we estimate the projected sensitivity in the  $\text{monojet} + \slashed E_\mathrm{T}$ channel at a 100~TeV $pp$ collider based on simulation.
The obtained results would be applicable to both the SPPC and FCC-hh projects.

In the simulation with $\sqrt{s}  = 100~\si{TeV}$, we consider the signal processes and only the primary SM backgrounds $W\left(\rightarrow \ell\nu\right)+\text{jets}$ and $Z\left(\rightarrow \nu\bar{\nu}\right)+\text{jets}$.
Other backgrounds should be small and can be safely neglected.
In the \texttt{Delphes} simulation, we conservatively assume that the future detector has the same parameters as those in the ATLAS detector.
The thresholds in the reconstruction and cut conditions are appropriately adjusted for a 100~TeV $pp$ collider, as also demonstrated in Table~\ref{t_monojet}.

Four signal regions are defined by requiring $\slashed{E}_\mathrm{T} > 1.5,~1.8,~2.2,~2.8~\si{TeV}$.
In each signal region, the signal significance $\mathcal{S}$ is defined as 
\begin{equation}\label{signif}
\mathcal{S}=\frac{S}{\sqrt{S+B}},
\end{equation}
where $S$ and $B$ are the estimated numbers of the signal events and the total background events passing the corresponding cuts, respectively.

\begin{table}[!t]
\centering
\caption{Information of the four benchmark points with the fixed coupling relation $\lambda_1 = \lambda_2 = 3\lambda_0/2$ for the $\text{monojet} + \slashed E_\mathrm{T}$ channel.}
\label{tab:monobmppoints}
\setlength{\tabcolsep}{.4em}
\renewcommand{\arraystretch}{1.2}
\begin{tabular}{cccccccc}
\hline\hline
&$M_X/\si{GeV}$ &$\lambda_2$&$m_{++}/\si{GeV}$ &$m_{\phi}/\si{GeV}$ &$m_{a}/\si{GeV}$ &$m_1/\si{GeV}$ &$m_2/\si{GeV}$  \\
\hline
BMP-a&400&0.2&405.0&400&419.7&401.0&418.7\\
BMP-b&700&0.2&702.9&700&711.4&700.6&710.9\\
BMP-c&1000&0.2&1002.0&1000&1008.0&1000.4&1007.6\\ 
BMP-d&1300&0.2&1301.6&1300&1306.2&1300.3&1305.9\\
\hline\hline
\end{tabular}
\end{table}

\begin{table}[!t]
\centering
\caption{Visible cross section $\sigma_\mathrm{vis}$ in femtobarns and signal significance $\mathcal{S}$ for integrated luminosity 3~$\si{ab^{-1}}$ after each cut in the signal region with $\slashed E_\mathrm{T}> 1.5~\si{TeV}$ of the $\text{monojet} + \slashed E_\mathrm{T}$ channel at $\sqrt{s} = 100~\si{TeV}$.}
\label{tab:monocutres_xsec}
\setlength{\tabcolsep}{.33em}
\renewcommand{\arraystretch}{1.2}
\begin{tabular}{ccccccccccccc}
\hline\hline
&   $W\to \ell\nu$&   $Z\to\nu\bar{\nu}$ &\multicolumn{2}{c}{BMP-a}&\multicolumn{2}{c}{BMP-b}&\multicolumn{2}{c}{BMP-c}&\multicolumn{2}{c}{BMP-d}  \\ 
&  $\sigma_\mathrm{vis}$&$\sigma_\mathrm{vis}$ &$\sigma_\mathrm{vis}$&$\mathcal{S}$ &$\sigma_\mathrm{vis}$&$\mathcal{S}$ &$\sigma_\mathrm{vis}$&$\mathcal{S}$ &$\sigma_\mathrm{vis}$&$\mathcal{S}$\\
\hline  
Cut~1&   6080 &  1481 &  8.08 & 5.08 & 3.16 & 1.99 & 1.41& 0.89 &0.73 & 0.46\\
Cut~2&   4428 &  1481 &  7.79 & 5.54 & 3.11 & 2.21 & 1.41& 1.00 &0.73 & 0.52\\
Cut~3&   1442 &  654  &  5.24 & 6.26 & 2.32 & 2.77 & 1.07 & 1.27 &0.56  & 0.66\\
Cut~4&   62.7   &  139  &  3.65 & 13.8 & 1.80 & 6.81 & 0.87 & 3.31 &0.47  & 1.79 \\
\hline
\hline
\end{tabular}
\end{table}

In order to show the cut efficiency in the $\text{monojet} + \slashed E_\mathrm{T}$ channel, we adopt four benchmark points (BMPs) for the QSDM model, whose parameters and mass spectra are listed in Table~\ref{tab:monobmppoints}.
All of them satisfy $\lambda_1=\lambda_2=3\lambda/2$, leading to vanishing $h\phi\phi$ coupling and $m_{\phi}=m_X$.
Thus, they would not be constrained by direct detection.
We choose the same $\lambda_2$ but different $m_{X}$ for the four BMPs.
As discussed above, a larger $m_{X}$ leads to a more compressed mass spectrum.
The predicted relic abundances of these BMPs are lower than the observed value.

For the signal region with $\slashed E_\mathrm{T}> 1.5~\si{TeV}$, we divide the cut conditions into the following four cuts.
\begin{itemize}
\item \textit{Cut~1}.---At least one reconstructed jet, and the leading jet with $p_{\mathrm{T}}>1.4~\mathrm{TeV}$ and $|\eta|<2.4$.
\item \textit{Cut~2}.---No reconstructed lepton.
\item \textit{Cut~3}.---At most four reconstructed jets, and $\Delta\phi(j_i,\slashed{\mathbf{p}}_\mathrm{T})>0.4$.
\item \textit{Cut~4}.---$\slashed E_\mathrm{T}>1.5~\mathrm{TeV}$.
\end{itemize}
After applying the cuts one by one, the visible cross section $\sigma_\mathrm{vis}$ and the signal significance $\mathcal{S}$ for an integrated luminosity of 3~$\si{ab^{-1}}$ are tabulated in Table~\ref{tab:monocutres_xsec}.
While cut~2 does not affect the $Z\left(\rightarrow \nu\bar{\nu}\right)+\text{jets}$ background, it reduces the $W\left(\rightarrow \ell\nu\right)+\text{jets}$ background which has a genuine lepton in the final state. 
Cut~3 and cut~4 combined suppress the $W\left(\rightarrow \ell\nu\right)+\text{jets}$ [$Z\left(\rightarrow \nu\bar{\nu}\right)+\text{jets}$] background by 2 (1) orders of magnitude.
For all the BMPs, these cuts subsequently increase the signal significance.
Note that Eq.~\eqref{signif} does not take into account systematic uncertainties, which could be a few to ten percent in monojet searches.
If systematic uncertainties are considered, the signal significance would be reduced.
Since BMP-b, -c, and -d have a $S/B$ ratio below 1\%, it could be difficult to test them in this signal region.

Combining the four signal regions, the expected exclusion limits at 95\% C.L. are shown in Figs.~\ref{monoj0hxx} and \ref{monojjian0hxx}.
For datasets of $300~\si{fb^{-1}}$ and $3~\si{ab^{-1}}$ at $\sqrt{s} = 100~\si{TeV}$, monojet searches are expected to probe the DM candidate mass $m_\phi$ up to $\sim 700~\si{GeV}$ and $\sim 1.2~\si{TeV}$, respectively.
Thus, a 100~TeV $pp$ collider looks much more powerful than the LHC.

In order to compare with direct detection and relic abundance observation, we have also plotted the 95\% C.L. expected exclusion limits in the $\text{monojet} + \slashed E_\mathrm{T}$ channel at $\sqrt{s}=100~\si{TeV}$ with an integrated luminosity of $3~\si{ab^{-1}}$ in Figs.~\ref{DD_RD:lamb2} and \ref{DD_RD:lamb1}.
We find that the 100~TeV monojet searches could cover some regions where direct detection experiments cannot probe.
Nonetheless, the regions predicting an observed relic abundance could not be reached.

\section{Soft-lepton searches at $pp$ colliders}
\label{sec:soft}

Besides the monojet channel, leptons arising from the scalar decays $\chi_i\to \chi_j + W^{\pm(*)}(\to \ell^\pm \nu_\ell)/Z^{(*)}(\to \ell^\pm \ell^\mp)$ may also contain important information for exploring the QSDM model.
Inspired by the searches for electroweak production of charginos and neutralinos in supersymmetric models, we first consider the final states involving two or three ``hard'' leptons.
After recasting the related ATLAS analysis at $\sqrt{s} = 13~\si{TeV}$ with a dataset of $36.1~\si{fb^{-1}}$~\cite{Aaboud:2018jiw}, however, we do not find any meaningful constraint on the QSDM model.
The main reason is that the leptons from the scalar decays tend to be rather soft, because the mass spectrum is typically compressed, as explained in the previous section.

Therefore, it is more suitable to consider the final states with ``soft'' leptons.
In this case, a pair of same-flavor opposite-sign (SFOS) soft leptons with an invariant mass $\lesssim 60~\si{GeV}$ could lead to a distinct signature~\cite{Han:2014kaa,Baer:2014kya}.
In the signal process $pp\to \chi_i\chi_j +\text{jets}$, such a SFOS lepton pair may come from the scalar decays into an off-shell $Z$ boson.
In order to induce a sufficiently large $\slashed E_\mathrm{T}$, a hard jet with a transverse direction roughly opposite to that of $\slashed{\mathbf{p}}_\mathrm{T}$ is also required.
Such a $\text{soft-leptons} + \text{jets} + \slashed E_\mathrm{T}$ channel has been utilized in the ATLAS search for supersymmetric particles with compressed mass spectra at the 13~TeV LHC with an integrated luminosity of $36.1~\si{fb^{-1}}$~\cite{Aaboud:2017leg}.
Important SM backgrounds in this channel include $t\bar{t} + \text{jets}$, $tW + \text{jets}$, $VV + \text{jets}$, and $\tau^+\tau^- + \text{jets}$.

\subsection{LHC constraint}

\begin{table}[!t]
\caption{Reconstruction and cut conditions in the ATLAS $\text{soft-leptons} + \text{jets} + \slashed E_\mathrm{T}$ analysis at $\sqrt{s} = 13~\si{TeV}$~\cite{Aaboud:2017leg}.}
\setlength{\tabcolsep}{1em}
\renewcommand{\arraystretch}{1.2}
\label{t_soft}
\begin{tabular}{cc}
\hline\hline
\multicolumn{2}{c}{Reconstruction conditions}\\
Electron $p_\mathrm{T}$, $|\eta| $ & $>4.5~\mathrm{GeV}$, $<2.47$ \\
Muon $p_\mathrm{T}$, $|\eta| $ & $>4~\mathrm{GeV}$, $<2.5$ \\
Non-$b$-tagged jet $p_\mathrm{T}$, $|\eta| $ & $>30~\mathrm{GeV}$, $<2.8$ \\
$b$-tagged jet $p_\mathrm{T}$, $|\eta| $ & $>20~\mathrm{GeV}$, $<2.5$ \\
\hline
\multicolumn{2}{c}{Cut conditions}\\
Number of leptons &2\\
Lepton flavor and charge & $e^+e^-\ \mathrm{or}\ \mu^+\mu^-$\\
Leading lepton $p_\mathrm{T}^{\ell_1}$ & $>5~\mathrm{GeV}$ \\
Subleading lepton $p_\mathrm{T}^{\ell_2}$ & $>4.5\ (4)~\mathrm{GeV}$ for $\ell_2 = e~(\mu)$ \\
$\Delta R_{\ell\ell}$ & $0.05<\Delta R_{\ell\ell}<2$\\
$m_{\ell\ell}$ & $[1,3]\cup[3.2,60]~\mathrm{GeV}$ \\
$\slashed E_\mathrm{T}$ & $>200~\mathrm{GeV}$ \\
Number of jets & $\ge 1$\\
Leading jet $p_{\mathrm{T}}$ & $>100~\mathrm{GeV}$ \\
$\Delta \phi(j_1,\slashed{\mathbf{p}}_\mathrm{T})$ & $>2$ \\
$\mathrm{min}(\Delta\phi(j_i,\slashed{\mathbf{p}}_\mathrm{T}))$ & $>0.4$ \\
Number of $b$-tagged jets & $0$\\
$m_{\tau\tau}$ & $<0\ \mathrm{or}\ >160~\mathrm{GeV}$ \\
$m_{\mathrm{T}}^{\ell_1}$ & $<70~\mathrm{GeV}$ \\
$\slashed E_\mathrm{T}/H_{\mathrm{T}}^{\mathrm{lep}}$ &  $>\mathrm{max}(5,15-2m_{\ell\ell}/\mathrm{GeV})$\\
\hline\hline
\end{tabular}
\end{table}

We reinterpret the ATLAS analysis~\cite{Aaboud:2017leg} to study the current constraint on the QSDM model in the $\text{soft-leptons} + \text{jets} + \slashed E_\mathrm{T}$ channel.
The corresponding reconstruction and cut conditions are summarized in Table~\ref{t_soft}.
The $p_\mathrm{T}$ thresholds for reconstructed electrons and muons are lowered to $4.5$ and $4~\si{GeV}$ for keeping soft leptons in the final state.
There should be exact two leptons forming a SFOS pair,
whose direction distance $\Delta R_{\ell\ell}$ and invariant mass $m_{\ell\ell}$ should lie in proper ranges because they are considered to be originated from an off-shell $Z$ boson.
Events with $m_{\ell\ell} \in (3,3.2)~\si{GeV}$ are rejected to avoid contamination from $J/\psi$ decays.
In order to increase the signal-to-background ratio, at least one jet with $p_\mathrm{T}>100~\si{GeV}$ and $\slashed E_\mathrm{T} > 200~\si{GeV}$ are required.
The condition $\Delta \phi(j_1,\slashed{\mathbf{p}}_\mathrm{T}) > 2$ is used to ensure the transverse directions of the leading jet and $\slashed{\mathbf{p}}_\mathrm{T}$ are quite opposite.
In order to suppress the $t\bar{t} + \text{jets}$ and $tW + \text{jets}$ backgrounds, no $b$-tagged jet is allowed.

For further increasing the signal significance, some dedicated kinematic variables are utilized.
The $m_{\tau\tau}$ variable~\cite{Han:2014kaa,Baer:2014kya,Barr:2015eva,Aaboud:2017leg} constructed by the SFOS lepton pair is helpful for reducing the $\tau^+\tau^- + \text{jets}$ background.
The leading lepton transverse mass $m_\mathrm{T}^{\ell_1} = \sqrt{2(E_\mathrm{T}^{\ell_1} \slashed E_\mathrm{T} - \mathbf{p}_\mathrm{T}^{\ell_1} \cdot \slashed{\mathbf{p}}_\mathrm{T})}$ is required to satisfy $m_\mathrm{T}^{\ell_1} < 70~\si{GeV}$, in order to suppress the $t\bar{t} + \text{jets}$, $VV + \text{jets}$, and $W + \text{jets}$ backgrounds.
The ratio of $\slashed E_\mathrm{T}$ to the scalar sum of the transverse momenta of the leptons $H_{\mathrm{T}}^{\mathrm{lep}} = p_\mathrm{T}^{\ell_1} + p_\mathrm{T}^{\ell_2}$ is used to improve the signal-to-background discrimination for compressed spectra.

\begin{table}[!t]
\centering
\caption{Signal regions defined by the $m_{\ell\ell}$ bins in the ATLAS $\text{soft-leptons} + \text{jets} + \slashed E_\mathrm{T}$ analysis at $\sqrt{s} = 13~\si{TeV}$~\cite{Aaboud:2017leg}.}
\label{tab:soft_bin_xsec}
\setlength{\tabcolsep}{.4em}
\renewcommand{\arraystretch}{1.2}
\begin{tabular}{cccccccc}
\hline\hline
Signal regions & SR1 & SR2 & SR3 & SR4& SR5 & SR6 & SR7\\
\hline
$m_{\ell\ell}$ $(\mathrm{GeV})$& ~[1,3]~ &~[1,5]~ &~[1,10]~&~[1,20]~ &~[1,30]~  &~[1,40]~  &~[1,60]~\\
$\sigma_\mathrm{vis}^\mathrm{obs}$ $(\mathrm{fb})$&0.10 &0.18&0.34&0.61&0.59&0.72&0.80\\
\hline\hline
\end{tabular}
\end{table}

We find that seven signal regions in the ATLAS analysis~\cite{Aaboud:2017leg} could be sensitive to the QSDM model. They are defined with different inclusive $m_{\ell\ell}$ bins, as tabulated in Table~\ref{tab:soft_bin_xsec}.
Note that the $(3,3.2)~\si{GeV}$ interval has also removed in these bins.
We also list the corresponding 95\% C.L. observed limits on the visible cross section, $\sigma_\mathrm{vis}^\mathrm{obs}$.
We thus simulate signal samples and apply the above cuts to obtain 95\% C.L. exclusion limits on the QSDM model.

\begin{figure}[!t]
\centering
\subfigure[~$\lambda_2 \geq 0.1$~\label{0hxx1jsomo}]
{\includegraphics[width=.48\textwidth]{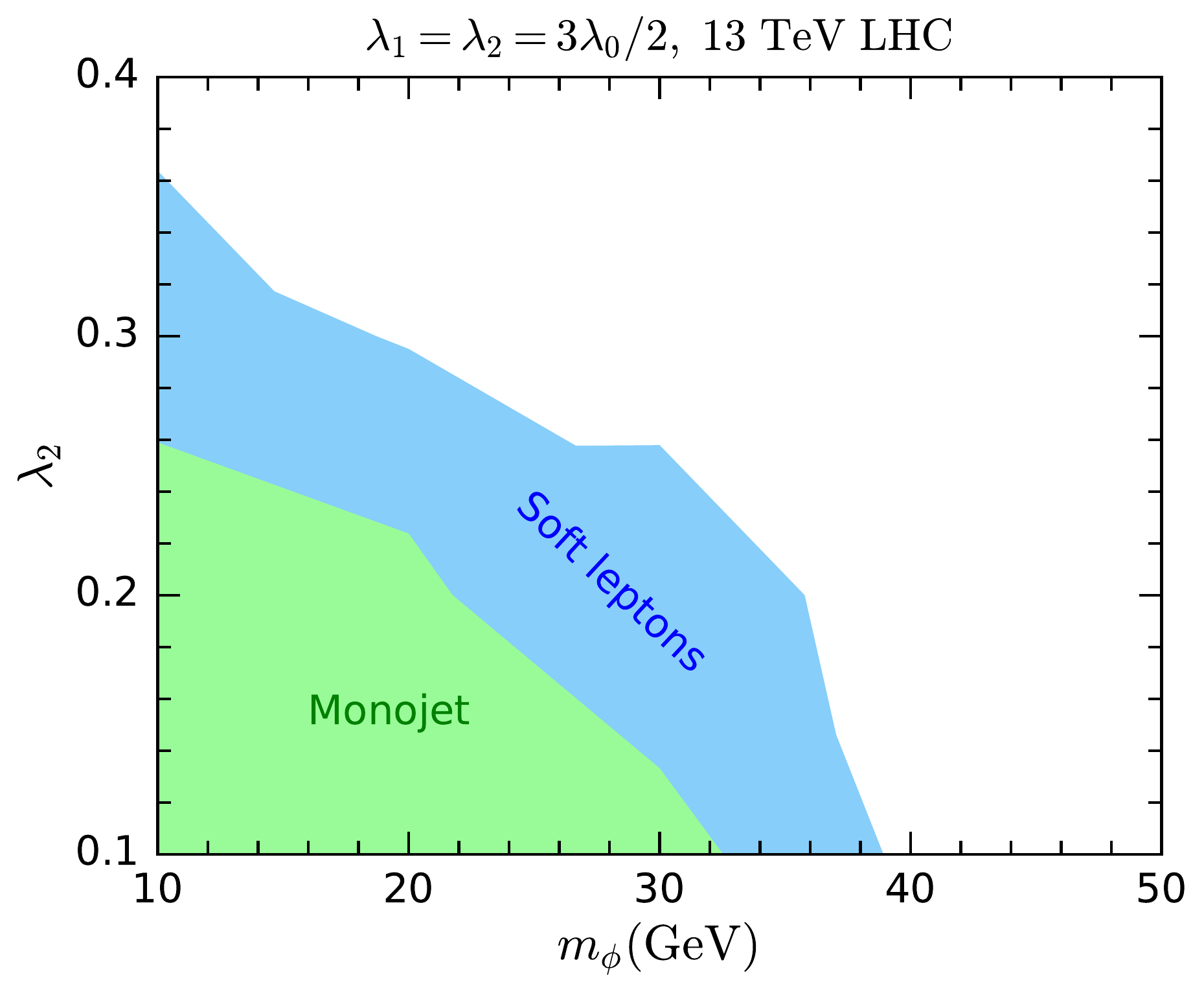}}
\subfigure[~$\lambda_2 \leq 0.09$ ~\label{small0hxxsomo}]
{\includegraphics[width=.48\textwidth]{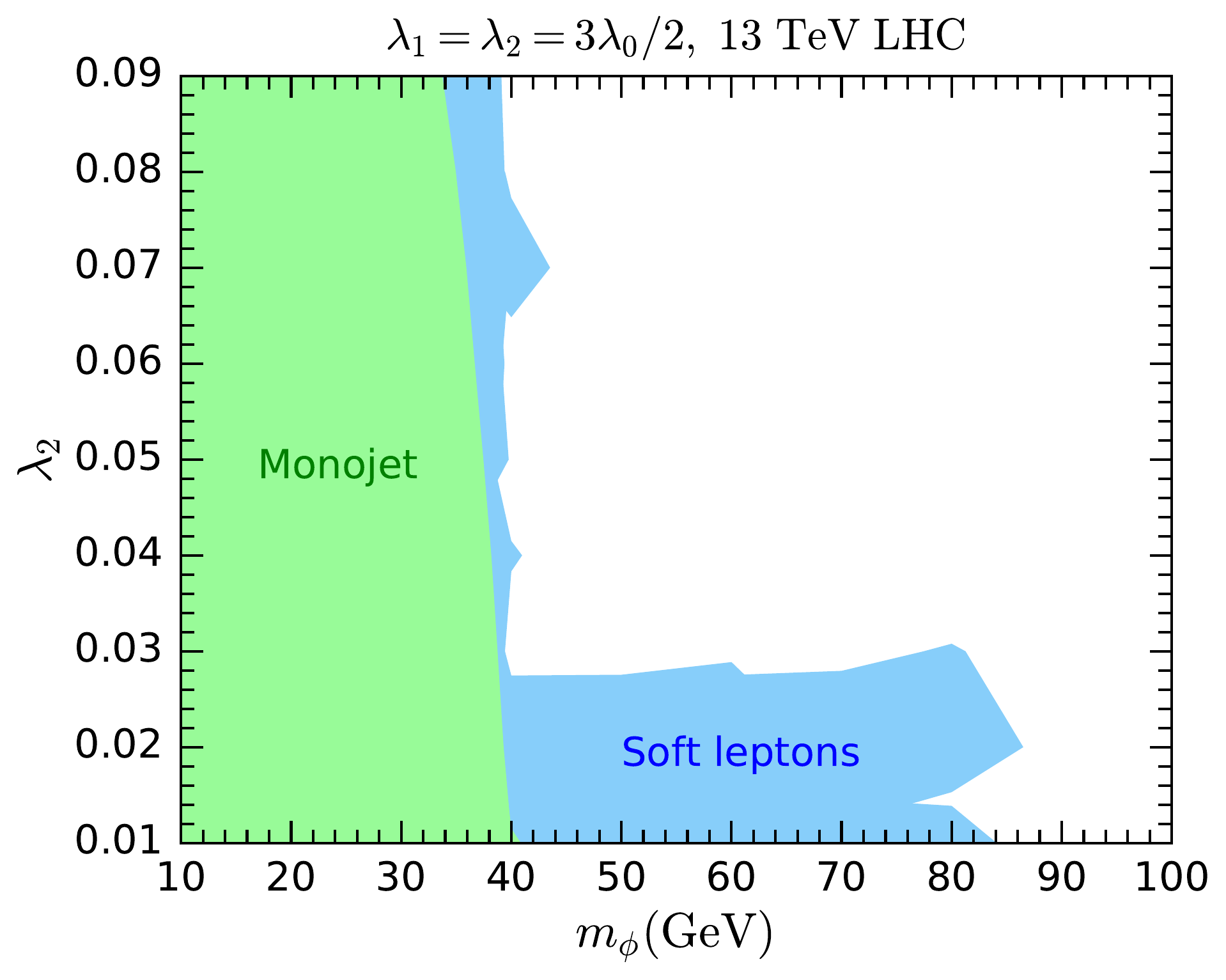}}
\caption{Constraints from current LHC searches in the $m_\phi$-$\lambda_2$ plane with the fixed coupling relation $\lambda_1=\lambda_2 = 3\lambda_0/2$ for $\lambda_2 \geq 0.1$ (a) and $\lambda_2 \leq 0.09$ (b).
Blue (green) regions are excluded at 95\% C.L. by the ATLAS $\text{soft-leptons} + \text{jets} + \slashed E_\mathrm{T}$~\cite{Aaboud:2017leg} ($\text{monojet} + \slashed E_\mathrm{T}$~\cite{Aaboud:2017phn}) analysis at the 13~TeV LHC with a dataset of $36.1~\si{fb^{-1}}$.}
\label{jian&0hxxsomo}
\end{figure}

The exclusion regions from the seven signal regions are combined, shown as the blue regions in Fig.~\ref{jian&0hxxsomo}.
The fixed coupling relation in Figs.~\ref{0hxx1jsomo} and \ref{small0hxxsomo} is identical to that in Fig.~\ref{monoj0hxx}.
For comparison, we also demonstrate the green regions excluded by the ATLAS monojet search, which has been discussed in the previous section.
We find that the $\text{soft-leptons} + \text{jets} + \slashed E_\mathrm{T}$ channel is more sensitive than the $\text{monojet} + \slashed E_\mathrm{T}$ channel at $\sqrt{s} = 13~\si{TeV}$.
In Fig.~\ref{0hxx1jsomo} for $\lambda_2 \geq 0.1$, the soft-lepton search has excluded a region with $m_\phi \lesssim 39~\si{GeV}$.
In Fig.~\ref{small0hxxsomo}, we focus on the small $\lambda_2$ region ($\lambda_2 \leq 0.09$) and find that the soft-lepton search can probe up to $m_\phi \sim 85~\si{GeV}$ for $\lambda_2 \lesssim 0.03$. 
The reason is that the $\slashed{E}_\mathrm{T}/H_{\mathrm{T}}^{\mathrm{lep}}$ cut is more suitable for small mass splittings, say, $m_a - m_\phi \lesssim 20~\si{GeV}$, which is realized in such a $\lambda_2 \lesssim 0.03$ region.

\subsection{Sensitivity at a 100 TeV $pp$ collider}

In this subsection, we explore the $\text{soft-leptons} + \text{jets} + \slashed E_\mathrm{T}$ channel at a 100~TeV $pp$ collider.
The main backgrounds $t\bar{t} + \text{jets}$, $tW + \text{jets}$, $VV + \text{jets}$, and $\tau^+\tau^- + \text{jets}$ are taken into account.
In order to demonstrate a detailed study, we choose four BMPs for this channel with the fixed coupling relation $\lambda_1 = \lambda_2 = 3\lambda_0/2$, which leads to $\lambda_{h\phi\phi} = 0$ and $m_\phi = m_X$.
The parameters and mass spectra of the BMPs are displayed in Table~\ref{tab:bmppoints}.
These BMPs would not be constrained by direct detection experiments, and they predict DM relic abundance lower than the observation.
BMP1 and BMP2 have identical $M_X$ and different $\lambda_2$, and, thus, a large $\lambda_2$ leads to larger mass splittings.
If $\lambda_2$ is fixed, a larger $M_X$ gives smaller mass splittings.
This can be seen by comparing BMP3 to BMP1 or BMP4 to BMP2.

\begin{table}[!t]
\centering
\caption{Information of the four benchmark points with the fixed coupling relation $\lambda_1 = \lambda_2 = 3\lambda_0/2$ for the $\text{soft-leptons} + \text{jets} + \slashed E_\mathrm{T}$ channel.}
\label{tab:bmppoints}
\setlength{\tabcolsep}{.4em}
\renewcommand{\arraystretch}{1.2}
\begin{tabular}{cccccccc}
\hline\hline
&$M_X/\si{GeV}$ &$\lambda_2$&$m_{++}/\si{GeV}$ &$m_{\phi}/\si{GeV}$ &$m_{a}/\si{GeV}$ &$m_1/\si{GeV}$ &$m_2/\si{GeV}$  \\
\hline
BMP1&400&0.2&405.0&400&419.7&401.0&418.7\\
BMP2&400&0.4&410.0&400&438.5&402.0&436.7\\
BMP3&500&0.2&504.0&500&515.9&500.8&515.1\\ 
BMP4&200&0.4&219.2&200&268.8&203.9&265.9\\
\hline\hline
\end{tabular}
\end{table}

For a $pp$ collider at $\sqrt{s} = 100~\si{TeV}$, we adopt the following reconstruction conditions with higher $p_\mathrm{T}$ thresholds than those used at the LHC.
\begin{itemize}
\item Reconstructed electrons are required to have $p_\mathrm{T} > 10~\mathrm{GeV}$ and $|\eta| <2.47$.
\item Reconstructed muons are required to have $p_\mathrm{T} > 10~\mathrm{GeV}$ and $|\eta| <2.5$.
\item Reconstructed non-$b$-tagged jets are required to have $p_\mathrm{T} > 60~\mathrm{GeV}$ and $|\eta| <2.8$.
\item Reconstructed $b$-tagged jets are required to have  $p_\mathrm{T} > 40~\mathrm{GeV}$ and $|\eta| <2.5$.
\end{itemize}
We appropriately modify the cut conditions according to a collision energy of 100~TeV.
They are classified into six subsequent cuts, as tabulated in Table~\ref{tab:sixCut}.
The $m_\mathrm{T}^{\ell_1}$ cut is abandoned, as we find that it would not be helpful.

\begin{table}[!t]
\centering
\caption{Cut conditions in the $\text{soft-leptons} + \text{jets} + \slashed E_\mathrm{T}$ channel at a 100~TeV $pp$ collider.}
\label{tab:sixCut}
\setlength{\tabcolsep}{1em}
\renewcommand{\arraystretch}{1.4}
\begin{tabular}{cc}
\hline\hline
Cut~1 & \tabincell{c}{Exact two 
SFOS leptons\\Leading lepton $p_{\mathrm{T}}>12~\mathrm{GeV}$, $0.05<\Delta R_{\ell\ell}<2$}\\
\hline
Cut~2 & \tabincell{c}{At least one jet, no $b$-tagged jet\\Leading jet $p_{\mathrm{T}}>200~\mathrm{GeV}$\\
$\Delta \phi(j_1,\slashed{\mathbf{p}}_\mathrm{T})>2.0$, $\mathrm{min}(\Delta\phi(j_i,\slashed{\mathbf{p}}_\mathrm{T})) >0.4$}\\
\hline
Cut~3 & $m_{\tau\tau}<0$ or $m_{\tau\tau}>200~\mathrm{GeV}$  \\
\hline
Cut~4 & $\slashed E_\mathrm{T}>280~\mathrm{GeV}$\\
\hline
Cut~5 & $\slashed E_\mathrm{T}/H_{\mathrm{T}}^{\mathrm{lep}}>\mathrm{max}(5,15-2m_{\ell\ell}/\mathrm{GeV})$ \\
\hline
Cut~6& $m_{\ell\ell}\in[1,3]\cup[3.2,60]~\mathrm{GeV}$ \\
\hline\hline
\end{tabular}
\end{table}

\begin{figure}[!t]
\centering
\subfigure[~Leading jet $p_\mathrm{T}$ distributions~\label{frac:j1pt}]
{\includegraphics[width=.48\textwidth]{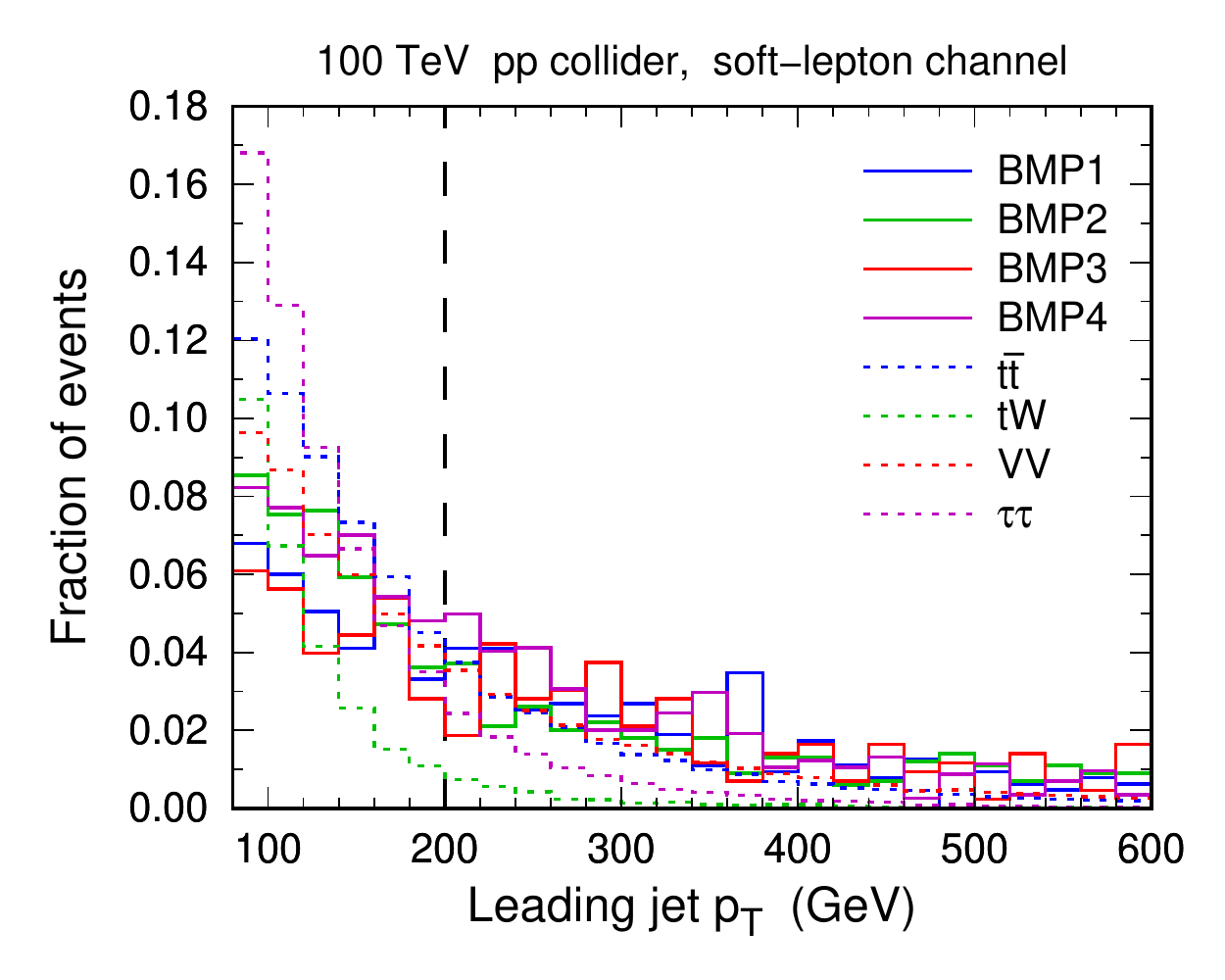}}
\subfigure[~$m_{\tau\tau}$ distributions~\label{frac:mtata}]
{\includegraphics[width=.48\textwidth]{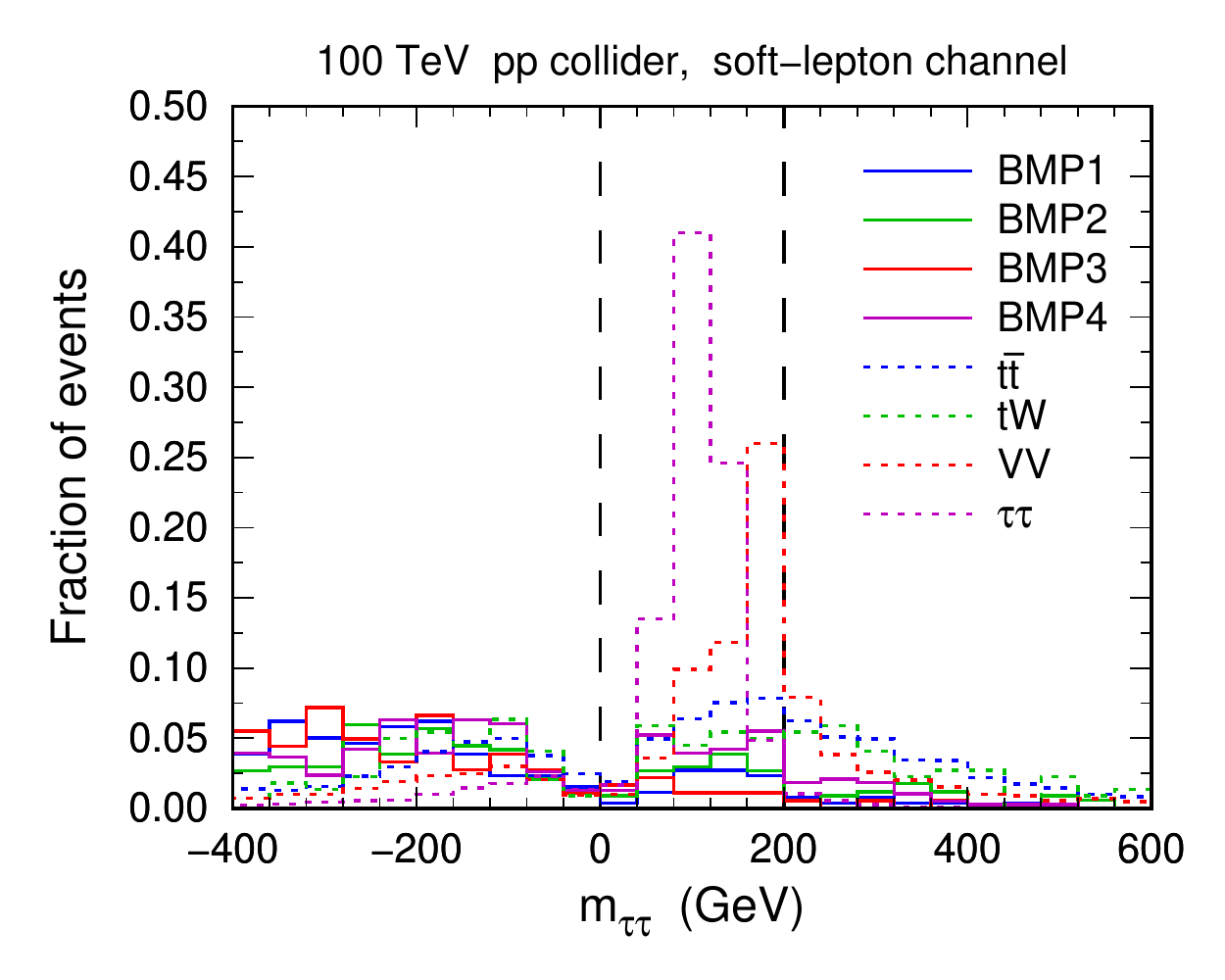}}
\caption{Fraction of signal and background events binned in the leading jet $p_\mathrm{T}$ after cut~1 (a) and in $m_{\tau\tau}$ after cut~2 (b) in the $\text{soft-leptons} + \text{jets} + \slashed E_\mathrm{T}$ channel at a 100~TeV $pp$ collider.
Dashed lines indicate the cut thresholds.}
\label{cutres1}
\end{figure}

Cut~1 selects the events with a proper soft SFOS lepton pair.
After applying cut~1, the fraction of events binned in the leading jet $p_\mathrm{T}$ for the four BMPs and for the backgrounds $t\bar{t} + \text{jets}$, $tW + \text{jets}$, $VV + \text{jets}$, and $\tau^+\tau^- + \text{jets}$ are presented in Fig.~\ref{frac:j1pt}.
We can see that these backgrounds tend to have lower  $p_\mathrm{T}$.
Thus, we require the leading jet $p_\mathrm{T}>200~\si{GeV}$ in cut~2 for reducing the backgrounds.

Figure~\ref{frac:mtata} shows the $m_{\tau\tau}$ distributions of signal and background events after cut~2.
The $m_{\tau\tau}$ variable is defined by $m_{\tau\tau} = \operatorname{sgn}(m_{\tau\tau}^2) \sqrt{|m_{\tau\tau}^2|}$
with $m_{\tau\tau}^2 \equiv (1+\xi_1)(1+\xi_2) m_{\ell\ell}^2$,
where $\xi_1$ and $\xi_2$ are parameters determined by solving $\slashed{\mathbf{p}}_\mathrm{T} = \xi_1 \mathbf{p}_\mathrm{T}^{\ell_1} + \xi_2 \mathbf{p}_\mathrm{T}^{\ell_2}$ event by event~\cite{Han:2014kaa,Baer:2014kya,Aaboud:2017leg,Barr:2015eva}.
If the $\tau$ leptons in the $pp\to Z^{(*)}/\gamma^{*} (\to \tau^+\tau^-)  + \text{jets}$ process both decay leptonically and the daughter neutrinos are collinear with the daughter charged leptons, such a $m_{\tau\tau}$ definition will truly correspond to the invariant mass of the $\tau$ leptons when the missing transverse momentum $\slashed{\mathbf{p}}_\mathrm{T}$ is genuinely contributed by the neutrinos.
Such a collinear situation would be realized when the two $\tau$ leptons are sufficiently boosted.
Consequently, the $m_{\tau\tau}$ distribution of the $\tau^+\tau^- + \text{jets}$ background peaks around $m_Z$, as demonstrated in Fig.~\ref{frac:mtata}.
Additionally, the $VV + \text{jets}$ distribution peaks around $2m_W$ because of the $W^+W^-\to \tau^+\tau^-\nu_{\tau}\bar\nu_{\tau}$ decay process.
Therefore, a veto on the events with $m_{\tau\tau} \in [0,200]~\si{GeV}$ in cut~3 can significantly suppress the $\tau^+\tau^- + \text{jets}$ and $VV + \text{jets}$ backgrounds.

\begin{figure}[!t]
\centering
\subfigure[~$\slashed E_\mathrm{T}$ distributions~\label{frac:mET}]
{\includegraphics[width=.48\textwidth,trim={0 6 0 10},clip]{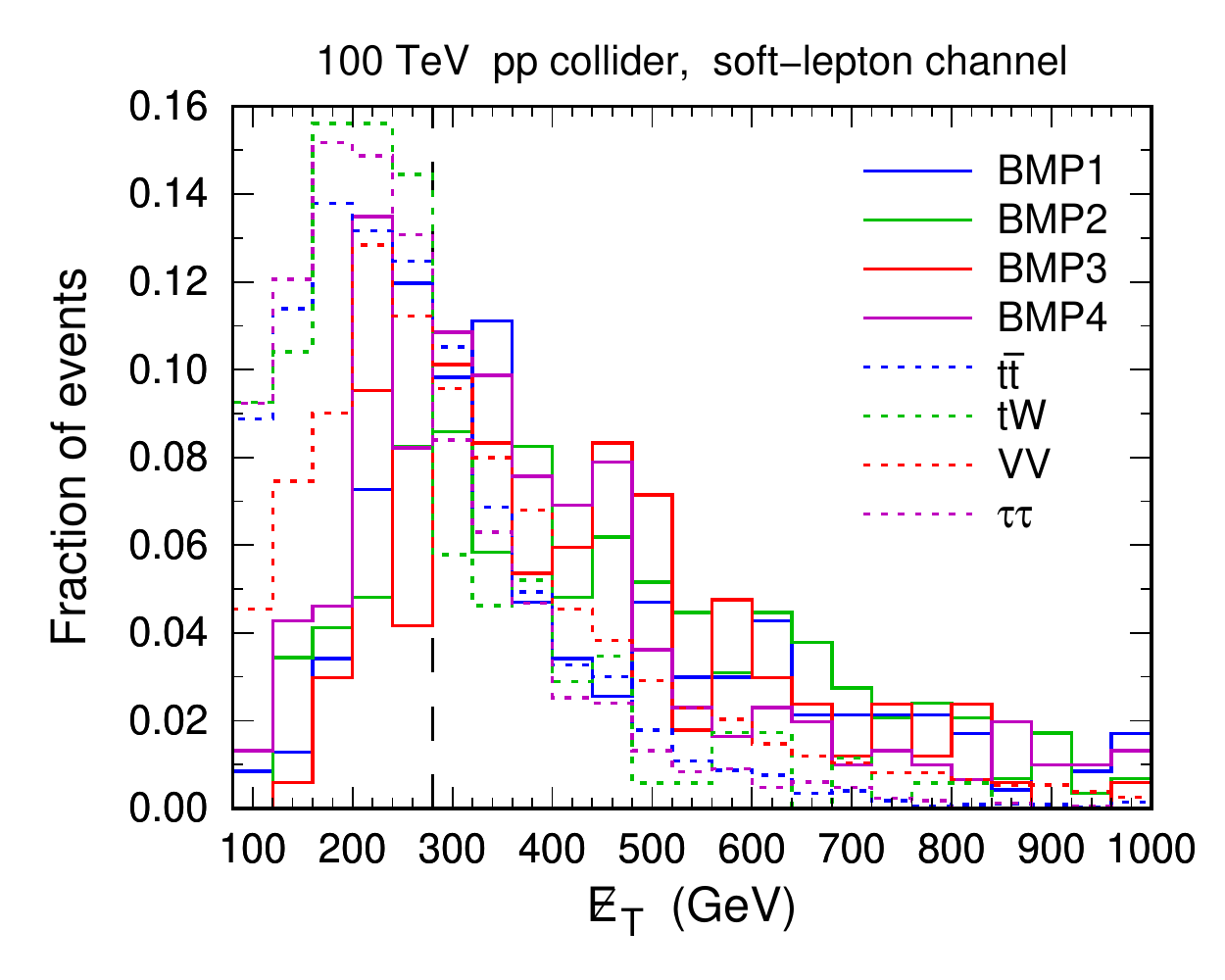}}
\subfigure[~$m_{\ell\ell}$ distributions~\label{frac:mll}]
{\includegraphics[width=.48\textwidth,trim={0 6 0 10},clip]{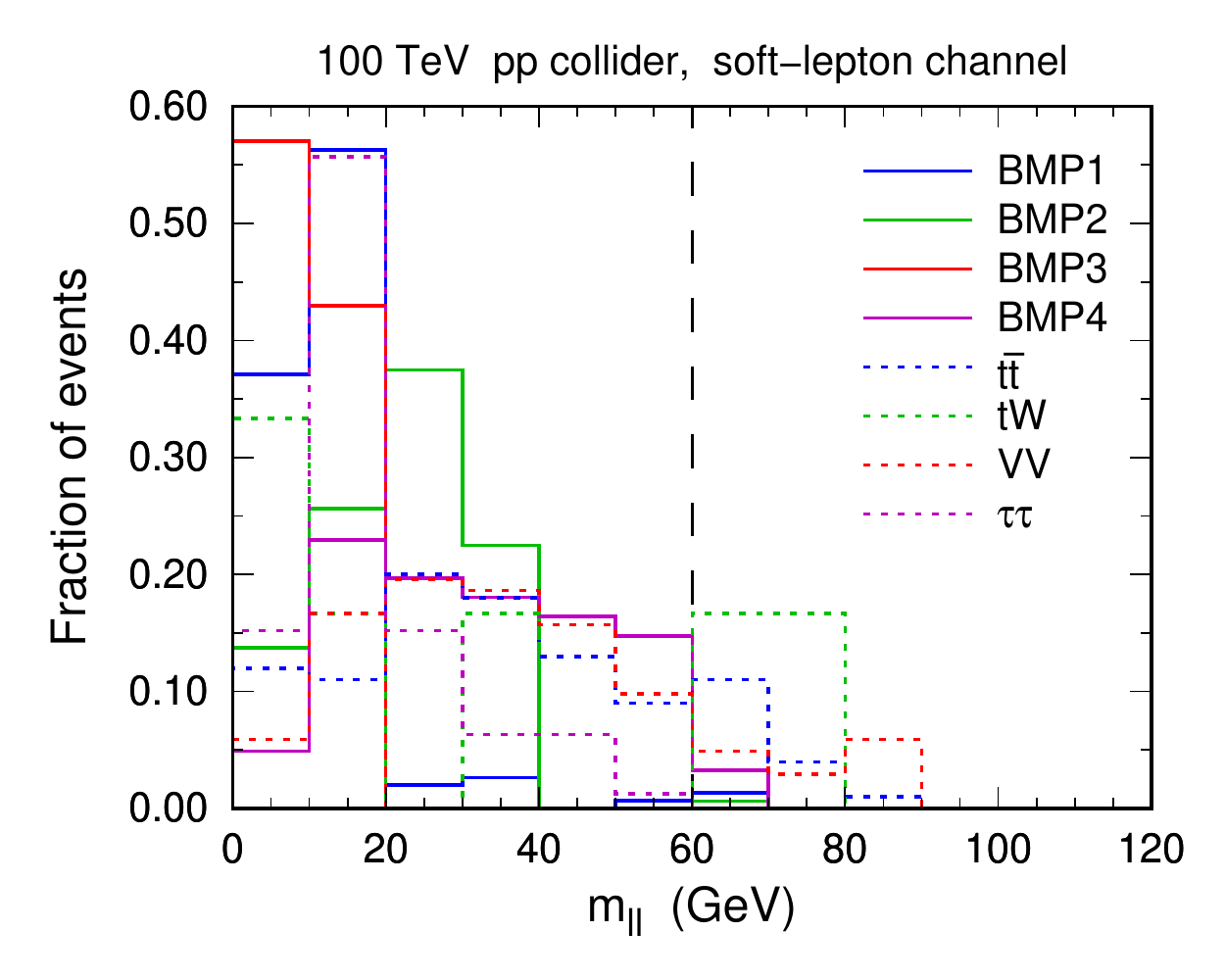}}
\caption{Fraction of signal and background events binned in $\slashed E_\mathrm{T}$ after cut~3 (a) and in $m_{\ell\ell}$ after cut~5 (b) in the $\text{soft-leptons} + \text{jets} + \slashed E_\mathrm{T}$ channel at a 100~TeV $pp$ collider.
Dashed lines indicate the cut thresholds.}
\label{cutres2}
\end{figure}

Figure~\ref{frac:mET} presents the $\slashed E_\mathrm{T}$ distributions after applying cut~3.
We find that the signal distributions are typically harder than the backgrounds, because the DM candidate $\phi$ with a mass of $\mathcal{O}(10^2)~\si{GeV}$ induces larger $\slashed E_\mathrm{T}$ than neutrinos.
Thus, we adopt the condition $\slashed E_\mathrm{T}>280~\mathrm{GeV}$ in cut~4 to increase the signal significance.
Cut~5 and cut~6 make use of the $\slashed E_\mathrm{T}/H_{\mathrm{T}}^{\mathrm{lep}}$ and $m_{\ell\ell}$ variables, following the ATLAS analysis~\cite{Aaboud:2017leg}.

The $m_{\ell\ell}$ distributions after applying cut~5 are displayed in Fig.~\ref{frac:mll}.
Inferring from Table~\ref{tab:bmppoints}, we have $m_a - m_\phi \sim 20,~ 39,~ 16,$ and $69~\si{GeV}$ for BMP1, BMP2, BMP3, and BMP4, respectively.
Such a difference in the mass splitting results in different end points in the $m_{\ell\ell}$ distributions, as clearly shown in Fig.~\ref{frac:mll}.
Seven signal regions are defined by the $m_{\ell\ell}$ bins as the same as those in Table~\ref{tab:soft_bin_xsec}.
Different $m_{\ell\ell}$ bins would be suitable for different mass splittings.

Table~\ref{tab:cutres_xsec} lists the visible cross section and the signal significance for an integrated luminosity of 3~$\si{ab^{-1}}$ after applying each cut in SR3.
We can see that the signal significances of the four BMPs subsequently increase from cut~1 to cut~5.
The cut condition $m_{\ell\ell} \in  [1,3]\cup[3.2,10]~\mathrm{GeV}$ in SR3 increases the  significances of BMP1 and BMP3 but decreases those of BMP2 and BMP4.
This is because BMP1 and BMP3 have smaller mass splittings and, hence, sufficient fractions of events satisfying $m_{\ell\ell} \leq 10~\si{GeV}$, while BMP2 and BMP4 do not, as shown in Fig.~\ref{frac:mll}.
Larger $m_{\ell\ell}$ bins in SR6 and SR7 would be applicable for BMP2 and BMP4.

\begin{table}[!t]
\centering
\caption{Visible cross section $\sigma_\mathrm{vis}$ in femtobarns and signal significance $\mathcal{S}$ for integrated luminosity 3~$\si{ab^{-1}}$ after each cut in SR3 of the $\text{soft-leptons} + \text{jets} + \slashed E_\mathrm{T}$ channel at $\sqrt{s} = 100~\si{TeV}$.}
\label{tab:cutres_xsec}
\setlength{\tabcolsep}{.33em}
\renewcommand{\arraystretch}{1.2}
\begin{tabular}{ccccccccccccc}
\hline\hline
&   $t\bar{t}$& $tW$& $VV$&   $\tau\tau$ &\multicolumn{2}{c}{BMP1}&\multicolumn{2}{c}{BMP2}&\multicolumn{2}{c}{BMP3}&\multicolumn{2}{c}{BMP4}  \\ 
&  $\sigma_\mathrm{vis}$  &$\sigma_\mathrm{vis}$&$\sigma_\mathrm{vis}$&$\sigma_\mathrm{vis}$ &$\sigma_\mathrm{vis}$&$\mathcal{S}$ &$\sigma_\mathrm{vis}$&$\mathcal{S}$ &$\sigma_\mathrm{vis}$&$\mathcal{S}$ &$\sigma_\mathrm{vis}$&$\mathcal{S}$\\
\hline  
Cut~1&  37600 &  28400&  5070 &  5420&  1.53 & 0.303& 2.29  & 0.453 &0.618 & 0.122 &17.3 & 3.43   \\
Cut~2&  1790  &  296  &  804  &  510 &  0.625& 0.586& 0.770 & 0.722 &0.262 & 0.246 &5.77 & 5.42   \\
Cut~3&  1280  &  232  &  383  &  73.3&  0.567& 0.699& 0.669 & 0.825 &0.243 & 0.300 &4.61 & 5.68\\
Cut~4&  445   &  69.7 &  190  &  21.7&  0.426& 0.863& 0.531 & 1.08  &0.201 & 0.408 &3.14 & 6.35\\
Cut~5&  37.3  &  8.04 &  9.91 &  3.47&  0.366& 2.57 & 0.368 & 2.59  &0.185 & 1.30  &0.934& 6.50\\
SR3&  4.11  &  2.68 &  0.583& 0.528&  0.136& 2.59 & 0.0483& 0.921 &0.106 & 2.02  &0.0455&0.868\\
\hline
\hline
\end{tabular}
\end{table}

\begin{figure}[!t]
\centering
\includegraphics[width=0.5\textwidth]{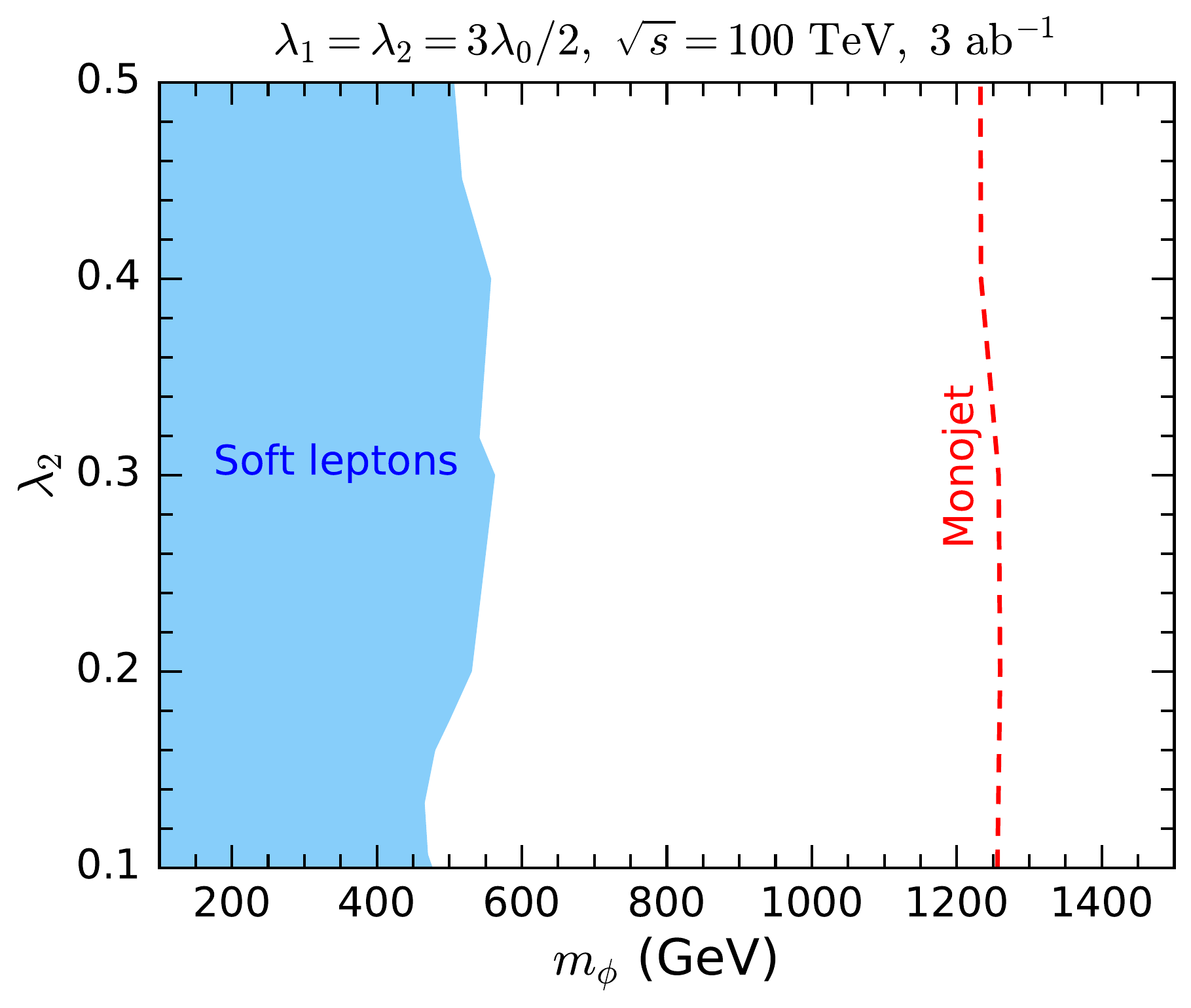}
\caption{95\% C.L. expected exclusion region in the $\text{soft-leptons} + \text{jets} + \slashed E_\mathrm{T}$ channel at a 100~TeV $pp$ collider with an integrated luminosity of $3~\si{ab^{-1}}$ for the fixed coupling relation $\lambda_1=\lambda_2 = 3\lambda_0/2$.
For comparison, the red dashed line denotes the 95\% C.L. expected exclusion limit in the $\text{monojet} + \slashed E_\mathrm{T}$ channel with the same collision energy and integrated luminosity.}
\label{fig:1000hxx1jsomo}
\end{figure}

Figure~\ref{fig:1000hxx1jsomo} shows the 95\% C.L. expected exclusion region combing the seven signal regions at a 100~TeV $pp$ collider with a dataset of $3~\si{ab^{-1}}$ for the fixed coupling relation $\lambda_1=\lambda_2 = 3\lambda_0/2$.
We find that the $\text{soft-leptons} + \text{jets} + \slashed E_\mathrm{T}$ channel can explore a region up to $m_\phi \sim 550~\si{GeV}$.
Nonetheless, such a sensitivity is not better than that in the $\text{monojet} + \slashed E_\mathrm{T}$ channel, which is demonstrated by the red dashed line.

\section{Conclusions and discussions}
\label{sec:con}

In this paper, we discuss the QSDM model, where the dark sector contains an inert $\mathrm{SU}(2)_\mathrm{L}$ quadruplet scalar with $Y=1/2$.
After the electroweak symmetry breaking, there are one doubly charged scalar, two singly charged scalars, and two neutral scalars.
For $\lambda_2>0$, the lighter neutral scalar $\phi$ plays the role of DM particle.
We have identified the parameter regions that can predict an observed DM relic abundance.

As the DM candidate can interact with nucleons through the SM Higgs portal, direct detection experiments could be sensitive to this model.
We have investigated the constraints from the current experiment XENON1T as well as the sensitivity of the future LZ experiment.
Nonetheless, the $h\phi \phi$ coupling could vanish if the quartic couplings $\lambda_0$, $\lambda_1$, and $\lambda_2$ satisfy special relations, resulting in null signal in direct detection.
In this case, other types of DM search experiments would be essentially important.

Since the dark sector scalars carry electroweak charges, they could be directly produced in pairs at high-energy $pp$ colliders.
The mass splittings among the dark sector scalars are typically lower than $m_W$ and $m_Z$.
As a result, the sensitive search channels at the LHC include the $\text{monojet} + \slashed E_\mathrm{T}$ and $\text{soft-leptons} + \text{jets} + \slashed E_\mathrm{T}$ channels.
We have recast the ATLAS analyses in these two channels with $\sqrt{s} = 13~\si{TeV}$ and an integrated luminosity of $36.1~\si{fb^{-1}}$.
We have found that the monojet search has excluded some parameter regions up to $m_\phi \sim 45~\si{GeV}$,
while the soft-lepton channel has excluded larger regions up to $m_\phi \sim 85~\si{GeV}$.

As these LHC constraints on the QSDM model still seem rather weak, we have studied the prospect of a future 100~TeV $pp$ collider, either SPPC or FCC-hh.
We have found that the monojet channel could be sensitive to the model up to $m_\phi \sim 1.2~\si{TeV}$ assuming an integrated luminosity of $3~\si{ab^{-1}}$.
On the other hand, the soft-lepton channel is less sensitive, reaching up to $m_\phi \sim 550~\si{GeV}$.

Electroweak precision measurements provide an indirect probe to the QSDM model.
The future determination of electroweak oblique parameters in the CEPC project would be able to reach up to $m_\phi \sim 0.6\text{--}1.3~\si{TeV}$.
But a direct search in the monojet channel at a 100~TeV $pp$ collider seems more sensitive in most regions.

Compared to the IDM, the QSDM model involves more electroweakly interacting dark sector scalars living in a larger $\mathrm{SU}(2)_\mathrm{L}$ representation.
This effectively enhances the annihilation and coannihilation cross sections of the scalars in the early Universe.
As a result, higher mass scales ($\gtrsim 2\text{--}3~\si{TeV}$) are required to yield the observed DM relic abundance.
Another consequence is that the pair production rates of the scalars at $pp$ colliders significantly increase.
Therefore, the LHC and a $100~\si{TeV}$ $pp$ collider are able to probe higher mass scales in the QSDM model than in the IDM (cf. Refs.~\cite{Arhrib:2013ela,Datta:2016nfz,Belyaev:2016lok,Belyaev:2018ext}).

\begin{acknowledgments}

This work is supported in part by the National Natural Science Foundation of China under Grants No.~11805288, No.~11875327, and No.~11905300, the China Postdoctoral Science Foundation
under Grant No.~2018M643282, the Natural Science Foundation of Guangdong Province
under Grant No.~2016A030313313,
the Fundamental Research Funds for the Central Universities,
and the Sun Yat-Sen University Science Foundation.

\end{acknowledgments}

\appendix

\section{Electroweak gauge interactions of the quadruplet scalar}
\label{app:gauge_part}

The generators in the $\mathrm{SU}(2)_\mathrm{L}$ representation $\mathbf{4}$ are given by
\begin{eqnarray}
&& T^1 = \left( {\begin{array}{*{20}{c}}
   {} & {\sqrt 3 /2} & {} & {}  \\
   {\sqrt 3 /2} & {} & 1 & {}  \\
   {} & 1 & {} & {\sqrt 3 /2}  \\
   {} & {} & {\sqrt 3 /2} & {}  \\
 \end{array} } \right),\quad
T^2 = \left( {\begin{array}{*{20}{c}}
   {} & {  -\sqrt 3 i/2} & {} & {}  \\
   {\sqrt 3 i/2} & {} & {  -i} & {}  \\
   {} & i & {} & {  -\sqrt 3 i/2}  \\
   {} & {} & {\sqrt 3 i/2} & {}  \\
 \end{array} } \right),
\nonumber\\
&&\hspace*{11em}
T^3 = {\mathrm{diag}}\left( {\frac{3}{2},\frac{1}{2}, - \frac{1}{2}, - \frac{3}{2}} \right).
\end{eqnarray}
Utilizing these generators, we can expand the gauge interaction terms for the quadruplet scalar as
\begin{eqnarray}
{\mathcal{L}_{{\mathrm{gauge}}}} &=& g\left[ {\frac{{\sqrt 6 }}{2}W_\mu ^ + {{({X^{ +  + }})}^*}i\overleftrightarrow {{\partial ^\mu }}{X^ + } + \sqrt 2 W_\mu ^ + {{({X^ + })}^*}i\overleftrightarrow {{\partial ^\mu }}{X^0} + \frac{{\sqrt 6 }}{2}W_\mu ^ + {{({X^0})}^*}i\overleftrightarrow {{\partial ^\mu }}{X^ - } + \mathrm{H.c.}} \right]
\nonumber\\
&&  + e{A_\mu }\Big[2{({X^{ +  + }})^*}i\overleftrightarrow {{\partial ^\mu }}{X^{ +  + }} + {({X^ + })^*}i\overleftrightarrow {{\partial ^\mu }}{X^ + } - {({X^ - })^*}i\overleftrightarrow {{\partial ^\mu }}{X^ - }\Big]
\nonumber\\
&&  + \frac{g}{{2{c_{\mathrm{W}}}}}{Z_\mu }\Big[(3c_{\mathrm{W}}^2 - s_{\mathrm{W}}^2){({X^{ +  + }})^*}i\overleftrightarrow {{\partial ^\mu }}{X^{ +  + }} + (c_{\mathrm{W}}^2 - s_{\mathrm{W}}^2){({X^ + })^*}i\overleftrightarrow {{\partial ^\mu }}{X^ + }
\nonumber\\
&& \qquad\qquad~ + iai\overleftrightarrow {{\partial ^\mu }}\phi  - (3c_{\mathrm{W}}^2 + s_{\mathrm{W}}^2){({X^ - })^*}i\overleftrightarrow {{\partial ^\mu }}{X^ - }\Big]
\nonumber\\
&& + \frac{{{g^2}}}{2}W_\mu ^ + {W^{ - \mu }}\big[ {3|{X^{ +  + }}{|^2} + 7|{X^ + }{|^2} + 7|{X^0}{|^2} + 3|{X^ - }{|^2}} \big] 
\nonumber\\
&& + {g^2}\left[ {\sqrt 3 W_\mu ^ + {W^{ + \mu }}{{({X^{ +  + }})}^*}{X^0} + \sqrt 3 W_\mu ^ + {W^{ + \mu }}{{({X^ + })}^*}{X^ - } + \mathrm{H.c.}} \right]
\nonumber\\
&&  + {e^2}{A_\mu }{A^\mu }\big(4|{X^{ +  + }}{|^2} + |{X^ + }{|^2} + |{X^ - }{|^2}\big)
\nonumber\\
&& + \frac{{eg}}{{{c_{\mathrm{W}}}}}{A_\mu }{Z^\mu }\big[2(3c_{\mathrm{W}}^2 - s_{\mathrm{W}}^2)|{X^{ +  + }}{|^2} + (c_{\mathrm{W}}^2 - s_{\mathrm{W}}^2)|{X^ + }{|^2} + (3c_{\mathrm{W}}^2 + s_{\mathrm{W}}^2)|{X^ - }{|^2}\big]
\nonumber\\
&&  + \frac{{{g^2}}}{{4c_{\mathrm{W}}^2}}{Z_\mu }{Z^\mu }\big[{(3c_{\mathrm{W}}^2 - s_{\mathrm{W}}^2)^2}|{X^{ +  + }}{|^2} + {(c_{\mathrm{W}}^2 - s_{\mathrm{W}}^2)^2}|{X^ + }{|^2} + |{X^0}{|^2} + {(3c_{\mathrm{W}}^2 + s_{\mathrm{W}}^2)^2}|{X^ - }{|^2}\big]
\nonumber\\
&&  + \Bigg\{ \left[ {\frac{{3\sqrt 6 }}{2}eg{A^\mu } + \frac{{\sqrt 6 {g^2}(2c_{\mathrm{W}}^2 - s_{\mathrm{W}}^2)}}{{2{c_{\mathrm{W}}}}}{Z^\mu }} \right]W_\mu ^ + {{({X^{ +  + }})}^*}{X^ + }
\nonumber\\
&& \qquad + \left( {\sqrt 2 eg{A^\mu } - \frac{{\sqrt 2 {g^2}s_{\mathrm{W}}^2}}{{{c_{\mathrm{W}}}}}{Z^\mu }} \right)W_\mu ^ + {{({X^ + })}^*}{X^0}
\nonumber\\
&& \qquad - \left[ {\frac{{\sqrt 6 }}{2}eg{A^\mu } + \frac{{\sqrt 6 {g^2}(2c_{\mathrm{W}}^2 + s_{\mathrm{W}}^2)}}{{2{c_{\mathrm{W}}}}}{Z^\mu }} \right]W_\mu ^ + {{({X^0})}^*}{X^ - } + \mathrm{H.c.} \Bigg\}.
\end{eqnarray}
Here $c_\mathrm{W}\equiv \cos \theta_\mathrm{W}$ and $s_\mathrm{W}\equiv \sin \theta_\mathrm{W}$, where $\theta_\mathrm{W} = \tan^{-1} (g'/g)$ is the weak mixing angle.

\bibliographystyle{utphys}
\bibliography{ref}
\end{document}